\documentclass[acmtog, nonacm]{acmart}

\AtBeginDocument{%
  \fancypagestyle{plain}{%
    \fancyhf{} %
    \fancyfoot[L]{
    © Owner/Author 2025.
    This is the author's version of the work.
    It is posted here for your personal use.
    Not for redistribution.
    It has been accepted for publication at the \textit{14th DACH+ Conference on Energy Informatics} and will be published in the \textit{ACM SIGEnergy Energy Informatics Review}.
    The definitive Version of Record will be available via the ACM Digital Library at: 
    \url{https://energy.acm.org/eir/}.}%
\fancyfoot[R]{}}
  \providecommand\BibTeX{{%
    \normalfont B\kern-0.5em{\scshape i\kern-0.25em b}\kern-0.8em\TeX}}}
    
\let\oldmaketitle\maketitle
\renewcommand{\maketitle}{%
  \oldmaketitle%
  \thispagestyle{plain}%
  \pagestyle{plain}}

\usepackage{subcaption}

\usepackage[acronym]{glossaries}
\usepackage{comment}
\usepackage{booktabs} 
\usepackage{pdflscape} 
\usepackage{longtable} 
\usepackage{pifont} 
\newcommand{\xmark}{\ding{55}}%
\newcommand{\cmark}{\ding{51}}%
\usepackage{enumitem} 
\usepackage{eurosym}
\usepackage{arydshln}
\usepackage{threeparttable}
\usepackage{multirow}
\usepackage{placeins}

\usepackage{tikz}
\usetikzlibrary{decorations.pathreplacing,matrix,patterns,positioning}

\usepackage[subpreambles=true]{standalone} 

\makeglossaries
\newacronym{ev}{EV}{electric vehicle}
\newacronym{da}{DA}{day-ahead}
\newacronym{cid}{CID}{continuous intraday}
\newacronym{id}{ID}{intraday}
\newacronym{ida}{IDA}{Intraday Auction}
\newacronym{ae}{AE}{absolute error}
\newacronym{se}{SE}{squared error}
\newacronym{mape}{MAPE}{mean absolute percentage error}
\newacronym{mae}{MAE}{mean absolute error}
\newacronym{rmspe}{RMSPE}{root mean squared percentage error}
\newacronym{mse}{MSE}{mean squared error}
\newacronym{rmse}{RMSE}{root mean squared error}
\newacronym{mspe}{MSPE}{mean squared percentage error}
\newacronym{crps}{CRPS}{continuous ranked probability score}
\newacronym{rebap}{reBAP}{"regelzonenübergreifender einheitlicher Bilanzausgleichsenergiepreis" (cross-control area
uniform balancing energy price)}
\newacronym{epex}{EPEX}{European Power Exchange}
\newacronym{cropex}{CROPEX}{Croatian Power Exchange}
\newacronym{hupx}{HUPX}{Hungarian Power Exchange}
\newacronym{vre}{VRE}{variable renewable energy}
\newacronym{tso}{TSO}{transmission system operator}
\newacronym{vwap}{VWAP}{volume-weighted average price}
\newacronym{vwsd}{VWSD}{volume-weighted standard deviation}
\newacronym{entsoe}{ENTSO-E}{European Network of Transmission System Operators for Electricity}
\newacronym{hpc}{HPC}{high performance computing}
\newacronym{pdf}{PDF}{probability distribution function}
\newacronym{rt}{RT}{real-time}
\newacronym{rw}{RW}{rolling window}
\newacronym{nrv}{NRV}{Netzregelverbund}
\newacronym{epf}{EPF}{electricity price forecasting}
\newacronym{sidc}{SIDC}{Single Intraday Coupling}
\newacronym{cont}{CONT}{continuous}
\newacronym{sdat}{SDAT}{Same Delivery Area Trading}
\newacronym{xbid}{XBID}{Cross Border Intraday}
\newacronym{daa}{DAA}{Day-Ahead Auction}
\newacronym{pca}{PCA}{principal component analysis}
\newacronym{lasso}{LASSO}{least absolute shrinkage and selection operator}
\newacronym{xgboost}{XGBoost}{eXtreme Gradient Boosting}
\newacronym{lstm}{LSTM}{long short-term memory}
\newacronym{ai}{AI}{artificial intelligence}
\newacronym{ml}{ML}{machine learning}
\newacronym{spf}{SPF}{single price forecasting} 
\newacronym{mpf}{MPF}{multiple price forecasting} 
\newacronym{tf}{TF}{trajectory forecasting} 
\newacronym{qh}{QH}{quarter-hourly} 
\newacronym{h}{H}{hourly} 
\newacronym{dm}{DM}{Diebold-Mariano}
\newacronym{adf}{ADF}{Augmented Dickey–Fuller}
\newacronym{cv}{CV}{cross-validation}
\newacronym{prob}{PROB}{probabilistic}
\newacronym{det}{DET}{deterministic}
\newacronym{cet}{CET}{central European time}
\newacronym{class}{CLS}{classification}
\newacronym{reg}{REG}{regression}
\newacronym{dsr}{DSR}{design science research}
\newacronym{auroc}{AUROC}{area under the receiver operating characteristic curve}
\newacronym{pnl}{PnL}{profit \& loss}
\newacronym{pls}{PLS}{partial least squares}
\newacronym{lob}{LOB}{limit order book}
\newacronym{nn}{NN}{neural network}

\newacronym{period}{PER}{period}
\newacronym{approach}{APP}{approach}
\newacronym{product_col}{PROD}{product}
\newacronym{target}{TGT}{target}
\newacronym{neighbor}{NBR}{neighbor}

\newglossaryentry{idfull}{name=\ensuremath{\text{ID}_{\text{full}}},description={}}
\newglossaryentry{id1}{name=\ensuremath{\text{ID}_1},description={}}
\newglossaryentry{id3}{name=\ensuremath{\text{ID}_3},description={}}
\newglossaryentry{m_32_bl}{name=\ensuremath{\mathcal{M}_{3\rightarrow2}^{\text{BL}}},description={}}
\newglossaryentry{m_21_bl}{name=\ensuremath{\mathcal{M}_{2\rightarrow1}^{\text{BL}}},description={}}
\newglossaryentry{m_10.5_bl}{name=\ensuremath{\mathcal{M}_{1\rightarrow0.5}^{\text{BL}}},description={}}

\newglossaryentry{dlvrystart}{
	name = \ensuremath{s} ,
	description = delivery start
}
\newglossaryentry{prodlen}{
	name = \ensuremath{\ell} ,
	description = product length
}
\newglossaryentry{product}{
	name = \ensuremath{(s,\ell)} ,
	description = product starting delivery at $s$ with a length of $\ell$
}
\newglossaryentry{prodtrades}{
	name = \ensuremath{\mathcal{T}^{s,\ell}} ,
	description = {set of trades in product $(s,\ell)$}
}
\newglossaryentry{tradevol}{
	name = \ensuremath{V_k^{s,\ell}} ,
	description = {volume of trade $k$ in product $(s,\ell)$}
}
\newglossaryentry{tradeprice}{
	name = \ensuremath{P_k^{s,\ell}} ,
	description = {price of trade $k$ in product $(s,\ell)$}
}
\newglossaryentry{tradetime}{
	name = \ensuremath{E_k^{s,\ell}} ,
	description = {price of trade $k$ in product $(s,\ell)$}
}
\newglossaryentry{tradesubset}{
    name = \ensuremath{\mathcal{S}^{s,\ell}} ,
    description = {subset of trades trades in product $(s,\ell)$}
}
\newglossaryentry{fcsttimes}{
	name = \ensuremath{\mathcal{U}_\gamma^{s,\ell}} ,
	description = {set of forecasting times for product $(s,\ell)$ and horizon length $\gamma$}
}
\newglossaryentry{regtarget}{
	name = \ensuremath{R^{s,\ell}_{u,\gamma}} ,
	description = {regression target for product $(s,\ell)$ of horizon length $\gamma$ at forecasting time $u$}
}
\newglossaryentry{dirtarget}{
	name = \ensuremath{D^{s,\ell}_{u,\gamma}} ,
	description = direction target
}
\newglossaryentry{baseintvl}{
	name = \ensuremath{\mathbb{B}_{u,\omega}} ,
	description = {time interval $[u-\omega,u)$}
}
\newglossaryentry{tradesinintvl}{
	name = \ensuremath{\mathcal{S}^{s,\ell}_{u,\omega}} ,
	description = {set of trades in product $(s,\ell)$ during interval $[u-\omega,u)$}
}
\newglossaryentry{curtime}{
	name = \ensuremath{t} ,
	description = {current time}
}
\newglossaryentry{curprice}{
	name = \ensuremath{\text{VWAP}_{\text{cur}, \gls{curtime}}^{s,\ell}} ,
	description = {current price}
}
\newglossaryentry{lagprice}{
	name = \ensuremath{\text{VWAP}_{\gls{curtime},h,\Delta}^{s,\ell}} ,
	description = {lagged price average}
}
\newglossaryentry{lagpricesd}{
	name = \ensuremath{\text{VWSD}_{\gls{curtime},h,\Delta}^{s,\ell}} ,
	description = {current price standard deviation}
}
\newglossaryentry{priceftr}{
	name = \ensuremath{\mathbf{x}^{s,\ell}_{\text{price},t}} ,
	description = {current price}
}

\begin{document}

\title{Directional Price Forecasting in the Continuous Intraday Market under Consideration of Neighboring Products and Limit Order Books}

\author{Timothée Hornek}
\email{timothee.hornek@uni.lu}
\orcid{0000-0001-6676-6541}

\author{Sergio Potenciano Menci}
\email{sergio.potenciano-menci@uni.lu}
\orcid{0000-0002-9032-7183}

\author{Ivan Pavić}
\email{ivan.pavic@uni.lu}
\orcid{0000-0002-5249-0625}
\affiliation{%
 \institution{SnT - Interdisciplinary Center for Security, Reliability and Trust University of Luxembourg}
  \city{Kirchberg}
  \state{Luxembourg}
  \country{Luxembourg}
}

\begin{abstract}
    The increasing penetration of \acrlong{vre} and flexible demand technologies, such as electric vehicles and heat pumps, introduces significant uncertainty in power systems, resulting in greater imbalance—defined as the deviation between scheduled and actual supply or demand.
Short-term power markets, such as the European \acrlong{cid} market, play a critical role in mitigating these imbalances by enabling traders to adjust forecasts close to real time.
Due to the high volatility of the \acrlong{cid} market, traders increasingly rely on \acrlong{epf} to guide trading decisions and mitigate price risk. 
However most \acrlong{epf} approaches in the literature simplify the forecasting task.
They focus on single benchmark prices, neglecting intra-product price dynamics and price signals from the \acrlong{lob}.
They also underuse high-frequency and cross-product price data.

In turn, we propose a novel directional \acrlong{epf} method for hourly products in the European \acrlong{cid} market.
Our method incorporates short-term features from both hourly and quarter-hourly products and is evaluated using German \acrlong{epex} data from 2024–2025.
The results indicate that features derived from the \acrlong{lob} are the most influential exogenous variables.
In addition, features from neighboring products—especially those with delivery start times that overlap with the trading period of the target product—improve forecast accuracy.
Finally, our evaluation of the value captured by our \acrlong{epf} suggests that the proposed \acrlong{epf} method has the potential to generate profit when applied in trading strategies.

\end{abstract}



\maketitle

\glsresetall

\sloppy
\FloatBarrier

\section{Introduction}\label{sec:introduction}

Production from \gls{vre} assets is inherently uncertain and variable due to its reliance on weather conditions~\cite{koch_short-term_2019}.
At the same time, consumption patterns are becoming more unpredictable. Mainly driven by the increasing integration of \glspl{ev} and heat pumps~\cite{ayyadi2019optimal}.
Together, these uncertainties contribute to imbalances—deviations between forecasted and actual power production or consumption—which may necessitate the activation of costly balancing energy~\cite{koch_passive_2020}.

To mitigate such imbalances, power market participants in Europe rely on short-term markets.
Among these markets, the \gls{cid} market plays a key role by enabling trading up to just a few minutes before physical delivery~\cite{meeus_evolution_2020}.
The \gls{cid} market operates continuously, with trades occurring at irregular intervals.
Thousands of trades can occur for the same product, where a product is a tradable electricity unit with a specific duration (e.g., 15, 30, or 60 minutes).
Overlapping trading periods further complicate the trading.
These irregular intervals and overlapping trading periods create a multi-time series forecasting problem that traditional models cannot easily address.

To deal with this complexity, traders are increasingly on short-term \gls{epf} methods to predict \gls{cid} price movements.
These \glspl{epf} often inform algorithmic trading strategies~\cite{the_netherlands_authority_for_consumers_and_markets_algorithmic_2024}.
A common \gls{epf} approach in literature is to forecast a single market benchmark per product because such an approach simplifies the task and frames it as a standard time series forecasting problem~\cite{kath_value_2018,uniejewski_understanding_2019,narajewski_econometric_2020,marcjasz_beating_2020}.
However, the approach of forecasting a single market benchmark per product ignores intra-product price fluctuations, failing to capture short-term price dynamics, which is crucial for effective algorithmic trading.
Recent research attempts to address this limitation by modeling trades as time series~\cite{narajewski_ensemble_2020,scholz_towards_2021,hirsch_simulation-based_2024,hirsch_multivariate_2024,hornek_comparative_2024}.
These studies typically focus on price averages or distributions.
Despite the growing interest in \gls{cid} \gls{epf} methods, we consider that several research gaps remain.

First, many approaches overlook high-frequency data.
For example, few incorporate more granular data than the 5-minute price granularity used in~\cite{narajewski_ensemble_2020,hirsch_simulation-based_2024,hirsch_multivariate_2024}.
Yet, short-term features—such as the \gls{vwap} of the last four trades or features derived from the \gls{lob}—can have a positive impact on improving forecast accuracy~\cite{scholz_towards_2021,hornek_comparative_2024}.
Second, prior research suggests that prices of related products influence forecast outcomes~\cite{hirsch_simulation-based_2024}.
However, most studies focus on related hourly products~\cite{hirsch_multivariate_2024}.
The potential role of quarter-hourly product prices in forecasting hourly products remains largely unexplored.
Third, directional price forecasting—used in both algorithmic and manual trading—remains underexplored in the academic \gls{cid} \gls{epf} literature.
To address these shortcomings, our study makes three key contributions to the existing \gls{cid} \gls{epf} literature:
\begin{enumerate}
    \item \textbf{Introduction of a Directional \gls{cid} \gls{epf} Method:} This study is the first to apply a classification approach to \gls{epf} in the \gls{cid} market, aligning more effectively with the requirements of algorithmic trading by focusing on directional price signals rather than point forecasts.
    \item \textbf{Expanded Use of Neighboring Product Features:} Building on prior work, the study incorporates both \acrlong{h} (\acrshort{h}) and \acrlong{qh} (\acrshort{qh}) neighboring products to capture inter-product dependencies, thereby enhancing forecast accuracy.
    \item \textbf{Integration of \gls{lob}-Derived Features in Directional \gls{epf}:} The study is among the first to explore \gls{lob} features within the context of directional \gls{epf}, addressing an underexplored yet potentially valuable data source.
\end{enumerate}
Our method is specifically designed for trading in the European power market.
It integrates short-term price features from both quarter-hourly and hourly products, incorporating data from \glspl{lob} as well as fundamental variables.
We evaluate the individual contribution of each feature set to forecasting accuracy.
Using \gls{epex} market data for Germany from 2024 to 2025, we find that \glspl{lob} and neighboring product prices significantly enhance forecast accuracy, whereas fundamental variables provide limited added value.
Notably, the greatest improvements arise from products whose delivery starts during the trading period of the target product.

The remainder of this manuscript is structured as follows.
Section~\ref{sec:market_description} reviews short-term power markets in Europe to contextualize our research.
Section~\ref{sec:related_works_and_contributions} presents a narrative literature review and provides essential background on \gls{cid} \gls{epf} methods.
Section~\ref{sec:modeling} describes our research process and modeling.
Section~\ref{sec:results_discussion} discusses our results in detail.
Finally, Section~\ref{sec:conclusion} concludes the manuscript.
\section{Background: Short-term European Power Markets}\label{sec:market_description}

In Europe, short-term power markets are commonly referred to as "spot markets", even though, in reality, they are futures markets with short lead times before the delivery periods~\cite{ku_leuven_energy_institute_ei_2015}.
These spot markets enable market participants to trade power from about 36 hours to a few minutes before delivery~\cite{ku_leuven_energy_institute_ei_2015}.
In Germany, a representative example of other European markets~\cite{birkeland_research_2024}, market participants can trade products with both hourly and quarter-hourly delivery periods, with hourly products being the most liquid~\cite{market_coupling_steering_committee_sidc_2023}.

The European spot market operates through two trading modalities: auctions and continuous trading.
These modalities differ in the way market participants trade and in the underlying market clearing mechanisms.
In auctions, the market clears once at a single price per product based on all orders submitted by participants before a certain gate closure time.
In Europe, there are two main types of auction markets: the \gls{daa} and the \gls{ida}. 
The \gls{daa} is a daily auction that is standardized across Europe, occurring once a day with a gate closure time of 12 noon , one day before the delivery for products spanning from 12 midnight on the current day to 12 midnight the following day.
All times, both previous and subsequent, are in \gls{cet}.
The \gls{ida} auction also occurs daily, with a gate closure time of 3~PM, one day before delivery.
For each product in both the \gls{daa} and \gls{ida}, offer only one price.
Consequently, these auctions produce conventional time series consisting of unique (timestamp, price) tuples for each product.

In contrast, continuous trading allows for orders to be submitted and cleared continuously, resulting in the dynamic clearing of prices in a \gls{lob}.
An order—which can be either a bid (buy) or an ask (sell)—is submitted by a market participant to buy or sell a power product at a specified price and volume.
Each product has a gate closure time shortly before delivery, after which trading is no longer possible.
This gate closure time varies by country, typically ranging from five minutes to hour before the start of delivery~\cite{all_nemo_committee_single_2021}.
Orders are continuously cleared at a central exchange, such as \gls{epex}~\cite{zachmann_design_2023}, where bids and asks are matched if the submitted prices align.
Each matched bid-ask pair establishes its own transaction price.
If bids and asks are not matched, they are placed in a \gls{lob} and await new orders with which they may be matched, or they remain unmatched after gate closure.

Since 2021, German spot markets no longer operate independently but have been coupled with other European markets through a shared algorithm~\cite{all_nemo_committee_single_2021}.
This coupling involves the direct modeling of interconnection lines between individual markets.
Generally, each European country forms a single bidding zone, where no internal interconnection lines are modeled.
However, some countries have multiple bidding zones—such as Sweden and Italy—or two countries can share a single bidding zone—as seen in the Luxembourgish-German bidding zone (due to Luxembourg's small size, it is included within the larger German zone). 
Typically, each bidding zone is managed by one \gls{tso}.
However, in larger zones, multiple delivery areas may exist, each operated by a different \gls{tso}.
This is the case in the Luxembourgish-German bidding zone, where there are four delivery areas managed by four \glspl{tso}: 50Hertz Transmission, Amprion, TenneT TSO, and TransnetBW.

\begin{figure}[h!]
    \centering
    \begin{tikzpicture}[xscale=.55, yscale=0.7]
    \draw (0,0) -- (1.8,0);
    \draw[dotted] (1.8,0) -- (2.7,0);
    \draw (2.7,0) -- (4.3,0);
    \draw[dotted] (4.3,0) -- (5.2,0);
    \draw[-stealth] (5.2,0) -- (13.5,0);

    \foreach \x in {1,3.5,6,8.5,11,12.5}
        \draw (\x cm,3pt) -- (\x cm,-3pt);
    
    \draw (1,0) node[below=1pt, font=\small] {12 noon};
    \draw (3.5,0) node[below=1pt, font=\small] {3 PM};
    \draw (6,0) node[below=1pt, font=\small] {$s\!-\!1h$};
    \draw (8.5,0) node[below=1pt, font=\small] {$s\!-\!30min$};
    \draw (11,0) node[below=1pt, font=\small] {$s\!-\!5min$};
    \draw (12.5,0) node[below=1pt+1.85pt, font=\small] {$s$};
    
    \draw (1,0) node[font=\small, above=1pt] {DAA};
    \draw (4.75,0.25) node[font=\small, above=10pt] {SIDC};
    \draw (6,1.3) node[font=\small, above=10pt] {CONT};
    \draw (9.75,1.3) node[font=\small, above=10pt] {SDAT};
    \draw (2.25,-.75) node[below=5pt, font=\small] {day before $s$};

    \draw[thick, decorate, decoration={brace, amplitude=10pt}, yshift=0pt] (3.5,1.3) --  (8.5,1.3);

   \draw[thick, decorate, decoration={brace, amplitude=10pt}, yshift=0pt] (3.5,.25) --  (6,.25);
   
    \draw[thick, decorate, decoration={brace, amplitude=10pt}, yshift=0pt] (8.5,1.3) --  (11,1.3);

    \draw[thick, decorate, decoration={brace, amplitude=10pt}, yshift=5pt](4.75,-.75)  --  (0,-.75);
    
\end{tikzpicture}
    \caption{Trading timeline for hourly products in the German spot market.}
    \label{fig:trading_timeline_h}
\end{figure}
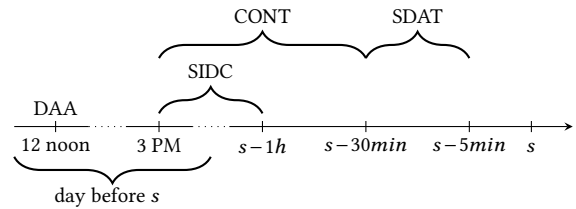

We illustrate the trading timeline for hourly products in the German power spot market in Figure~\ref{fig:trading_timeline_h}. 
The day before the delivery starts, denoted with ($s$), the \gls{daa} takes place at 12 midday (noon).
Following this, at 3~PM, \gls{cid} trading begins.
\Gls{cid} trading proceeds in various phases, defined by the respective bidding zones~\cite{hirth_handel_2021}.

During the \gls{cont} phase, products cover the entire German-Luxembourgish bidding zone, thus resulting in a single order book across this zone.
During the \gls{sidc} period, the German-Luxembourgish order book is shared with other European bidding zones, subject to limits in transmission capacities.
It is important to note that the \gls{cont} period exceeds the \gls{sidc} period.
During the portion of the \gls{cont} period not covered by the \gls{sidc}, trading with other bidding zones ceases, but trading continues with a single (German-Luxembourgish) order book. 
Subsequently, during the \gls{sdat} period, the trading splits into the four delivery areas within the German-Luxembourgish bidding zone, resulting in continuous trading with four distinct order books~\cite{hirth_handel_2021}.

Trading for quarter-hourly products follows a similar schedule, with slight modifications.
Since the \gls{daa} currently does not accommodate quarter-hourly products\footnote{Quarter-hourly products will be included in the \acrshort{daa} from 2025~\cite{epex_spot_se_new_2024}.}, the \gls{ida} is held at 3~PM on the day before delivery start ($s$) to enable these products to be traded.
Following the \gls{ida}, \gls{cid} trading for quarter-hourly products starts at 4~PM.

In summary, the \gls{daa} and \gls{cid} market are central to facilitating short-term power trading in the EU, but they use fundamentally different clearing mechanisms.
The \gls{daa} results in a single price per product, while the \gls{cid} involves multiple transactions per product, each with its own price.
Consequently, modeling and forecasting approaches differ significantly for each. 
The \gls{daa}'s single price per product allows for standard time series methods to forecast and model future cleared prices, whereas the \gls{cid}'s multiple transaction prices necessitates the use of more complex models to capture the full dynamics of price formation.

\section{Related Works and Contributions}\label{sec:related_works_and_contributions}

We conduct a narrative literature review following the guidelines of \cite{GREEN2006101} and ~\cite{pare_synthesizing_2015} with two main objectives: (1) to establish the current state of knowledge on \gls{epf} in the \gls{cid} market, and (2) to confirm the research gap we identified.
In the first step, we collect academic literature from major academic databases, including Google Scholar, IEEE Xplore, ACM Digital Library, and Scopus.
Our search strategy focuses on \gls{epf} methods specifically applied to European \gls{id} markets.
We perform keyword-based searches using terms such as \textit{"intraday electricity price forecasting"}, and \textit{"continuous intraday market"}.
In the second step, we evaluate the identified publications based on their titles, abstracts, and keywords.
Where necessary, we skim the full text to assess relevance.
We retain studies that focus on forecasting in \gls{id} or \gls{cid} markets, propose relevant methodologies, or offer insights into the structure and dynamics of \gls{id} trading.
This process allows us to synthesize methodological trends, identify recurring assumptions and limitations—such as the prevalent use of single benchmark prices or underutilization of high-frequency data—and clearly define the specific gap addressed by our proposed method.

We structure the remainder of this section as follows.
Section~\ref{sec:id_epf} reviews existing approaches to \gls{id} \gls{epf}.
Section~\ref{sec:general_studies} discusses broader studies related to the European \gls{id} market
Finally, Section~\ref{sec:review_summary} provides a summary of the narrative literature review findings.

\subsection{Intraday Electricity Price Forecasting}\label{sec:id_epf}

The existing body of \gls{id} \gls{epf} literature primarily concentrates on the German and Iberian (i.e., Spanish and Portuguese) \gls{id} markets~\cite{shinde_literature_2019}.
However, the structure of these markets differs considerably: the central European \gls{id} market is characterized by continuous trading, whereas the Iberian market operates primarily through multiple \gls{id} auctions.

In auction-based settings, each product clears at a single price, simplifying the formulation of \gls{epf} problems, as the resulting cleared prices generally follow conventional time series patterns.
In contrast, the continuous trading mechanism of the \gls{cid} market introduces significant challenges for \gls{epf}.
The occurrence of multiple transactions at irregular intervals for each product deviates from standard time series assumptions and complicates the modeling process.

Building on the classification proposed by~\citet{hirsch_multivariate_2024}, we identify three approaches in the literature for formulating \gls{epf} problems within the \gls{cid} context:
\begin{enumerate}[nosep]
    \item \textit{Single price forecasting} (see Section~\ref{sec:single_price_forecasting})
    \item \textit{Price trajectory forecasting} (see Section~\ref{sec:price_trajectory_forecasting})
    \item \textit{Multiple price forecasting} (see Section~ \ref{sec:multiple_price_forecasting})
\end{enumerate}

\subsubsection{Single Price Forecasting}\label{sec:single_price_forecasting}
In the single price forecasting approach, researchers forecast a single aggregate price for every product.
For instance \citet{kath_value_2018} forecasted the \gls{vwap} of all product transactions with an elastic net regression, and used their forecast to inform profitable trading strategies that sought arbitrage opportunities between the \gls{daa} and the \gls{cid} markets.
Similarly, \citet{uniejewski_understanding_2019} used a \gls{lasso} regression to predict \gls{id3} prices, which is a market index averaging all trades made between three hours to half an hour before the product delivery start.
The authors noted that using the prices of the most recent \gls{cid} transactions has a significant positive influence forecasting accuracy.
\citet{narajewski_econometric_2020} also focused on \gls{id3} index prices. They identified immediate prior \gls{cid} prices as being the most effective predictors for the movements of hourly products. They thereby claim that the market is "weak-form efficient," meaning that current market prices reflect all available information, and so future price changes cannot be explained by past prices.
Contrarily, \citet{marcjasz_beating_2020} challenge the weak form market efficiency hypothesis by showing empirically that the last price made before the \gls{id3} period is not the most effective prediction data point.
The \gls{id3} is a price index for the \gls{cid} market published by \gls{epex} covering trades during the three hours prior to delivery~\cite{epex_spot_se_description_2023}.
The authors trained a \gls{lasso} model that outperforms naive predictors.
\citet{maciejowska_pca_2020} introduced a \gls{pca}-based forecast averaging method to enhance \gls{id3} price forecasting.
More precisely, the authors averaged multiple forecasts using a \gls{pca} and find that the approach yields more precise forecasts than point forecasts or other forecast averaging techniques.

\citet{maciejowska_day-ahead_2019} followed a slightly modified approach by forecasting price spreads between the \gls{da} and \gls{cid} markets, and analyzing the economic implications of predicting price directionality. They highlighted that accuracy in sign classification does not guarantee the profitability of trading strategies.
\citet{cramer_multivariate_2023} proposed a probabilistic model to forecast the price differences between between \gls{da} and \gls{cid} markets.
They found that the most recent price history and increments in the \gls{da} market have the most influence on the price difference.

Overall, single price forecasting literature suggests that the \gls{cid} market is not weak form efficient, meaning that it is possible to forecast price movements from historical price data.
Nevertheless, Single Price Forecasting falls short as it does not account for changing prices within \gls{cid} trading sessions.

\subsubsection{Trajectory Forecasting}\label{sec:price_trajectory_forecasting}
In the trajectory forecasting approach, researchers predict an entire sequence of future prices—referred to as a price trajectory—to capture the expected development of prices over time.
However, the complete trajectory is forecasted at once, without incorporating new market information that may become available during the forecasting horizon.
\citet{serafin_trading_2022} used trajectory forecasting to inform a trading strategy that results in profits being generated.
This trading strategy involves timing the sale of power volumes produced by a renewable energy source on spot markets.
It outperformed several naive approaches, such as those that sell at the beginning or end of the trading period.
To our knowledge, these are the only authors who have studied price trajectory forecasting models based on our literature review.
Also, despite enabling the model to account for changing prices during trading sessions, the modeling approach does not account for changing market information between different forecasting steps made for single products.

\subsubsection{Multiple Price Forecasting}\label{sec:multiple_price_forecasting}
Multiple price forecasting considers each product as a separate time-series.
\citet{narajewski_ensemble_2020} employed ensemble forecasting techniques to predict the distribution of price trajectories, incorporating data that includes the introduction of \gls{xbid} in 2018.
Ensemble forecasting refers to a stochastic forecasting approach, involving the generation of multiple forecasts (i.e., the forecast ensemble).
They noted a decrease in market volatility after the introduction of \gls{xbid}.
Furthermore, \citet{hirsch_simulation-based_2024} adopted a similar approach to forecast the distribution of prices.
They concluded that fundamental variables, such as outages and \gls{vre} forecasts, do not impact \gls{cid} returns.
Instead, they found that volatility is primarily influenced by the merit-order regime, the time until delivery, and the time of the closure of cross-border order books.
In subsequent work, \citet{hirsch_multivariate_2024} expanded their modeling to consider cross-product effects using copulas.
They concluded, that considering cross-product dependencies improves forecasting accuracy.

Focusing on non-probabilistic price forecasting, \citet{scholz_towards_2021} adopted a \gls{rw} framework over $15min$ intervals.
By comparing different \gls{ml}/\gls{ai} models, they identified that the \gls{xgboost} algorithm excels in short-term forecasting (up to 30 minutes ahead), while the \gls{lstm} model excels for longer-term forecasts (up to one hour ahead).
Also using a \gls{rw} approach, \citet{hornek_comparative_2024} explored various baseline models that use the \glspl{vwap} of a set of the most recent transactions to forecast the \gls{vwap} for all trades in upcoming intervals, ranging from $1min$ to $60min$.
Their findings indicate that taking the \gls{vwap} of the last four trades exceeds the predictive performance of other baseline models they tested.

While the multiple price forecasting approach adds complexity to modeling, it is the only method that effectively captures market dynamics in \gls{cid} trading sessions.
This makes it potentially useful for informing trading decisions during \gls{cid} trading.

\subsection{General Studies}\label{sec:general_studies}
Other studies focused on explaining \gls{cid} prices rather than seeking to make direct price forecasts.
\citet{kiesel_econometric_2017} used an econometric model to show how prices asymmetrically adjust to \gls{vre} forecast errors.
Extending this, \citet{kremer_fundamental_2020} observed price reversion, emphasizing the role of price information in driving \gls{cid} prices and noting the influence of neighboring product prices.
They also found that \gls{vre} has a greater impact on prices in environments with a steep merit order.
Focusing on night-time trading, \citet{kremer_intraday_2020} highlighted the dominance of price information in night products, with supply and demand factors being less significant.

In terms of price volatility analysis and forecasting, \citet{kulakov_impact_2021} found that \gls{vre} forecast errors affect \gls{cid} price volatility in a non-linear fashion.
\citet{ziel_modeling_2017} detected no difference in the impact of positive versus negative \gls{vre} forecast errors on \gls{cid} prices.
\citet{baule_volatility_2021} identified a positive correlation between wind energy share, traded volume, and price deviations between \gls{da} and \gls{id} with regard to \gls{cid} price volatility.
To improve the accuracy of volatility forecasting, \citet{ciarreta_modeling_2017} evaluated volatility forecasting models, noting that their performance depends on the transformation applied to price data.
\citet{gurtler_effect_2018} investigated them impact of \gls{vre} generation on \gls{da} and \gls{cid} price volatility, suggesting that better \gls{vre} forecasting could reduce price volatility.

Research also extends to other markets, such as the Iberian \gls{id} market studied by \citet{frade_influence_2018}, who found that while wind power impacts volatility, it does not significantly change prices.
Additional analysis by \citet{andrade_probabilistic_2017} used quantile regression for probabilistic \gls{id} \gls{epf}, finding that large-scale generation outputs, system load, and \gls{vre} forecasts do not improve forecast quality.

\subsection{Summary}\label{sec:review_summary}
The literature on \gls{id} \gls{epf} indicates that the most recent transactions and prices of neighboring products significantly explain future \gls{id} price movements~\cite{hirsch_simulation-based_2024,kremer_fundamental_2020,kremer_intraday_2020}.
Additionally, fundamentals including \gls{vre} forecasting errors, production volumes, and the time of the closure of cross-border order books have a greater impact on price volatility than price movements~\cite{hirsch_simulation-based_2024,baule_volatility_2021,frade_influence_2018,andrade_probabilistic_2017}.
Moreover, the literature suggests that the \gls{cid} market is not weak-form efficient, and so they offer opportunities for forecasters to make useful forecasts of future price changes~\cite{marcjasz_beating_2020}.

\begin{table}[h!]
    \centering
    \begin{threeparttable}
    \caption{Overview of \gls{cid} \gls{epf} literature, documenting the test \gls{period}, forecasting \gls{approach}, forecasted \glspl{product_col}, and \gls{target} variable for each study.}
    \label{tab:epf_literature_overview}
    \begin{tabular}{lcccc}
    \toprule
    Study&\acrshort{period}&\acrshort{approach}&\acrshortpl{product_col}&\acrshort{target}\\
    \midrule
    \citet{kath_value_2018}&15-18&\acrshort{spf}&\acrshort{qh}&\gls{idfull}\\
    \citet{uniejewski_understanding_2019}&15-18&\acrshort{spf}&\acrshort{h}&\gls{id3}\\
    \citet{narajewski_econometric_2020}&15-18&\acrshort{spf}&\acrshort{qh}, \acrshort{h}&\gls{id3}\\
    \citet{marcjasz_beating_2020}&15-18&\acrshort{spf}&\acrshort{h}&\gls{id3}\\
    \citet{maciejowska_pca_2020}&15-19&\acrshort{spf}&\acrshort{h}&\gls{id3}\\
    \citet{maciejowska_day-ahead_2019}&16-17&\acrshort{spf}&\acrshort{h}&\gls{idfull}${}^*$\\
    \citet{cramer_multivariate_2023}&18-19&\acrshort{spf}&\acrshort{qh}&\gls{id3}${}^*$\\
    \citet{serafin_trading_2022}&17-19&\acrshort{tf}&\acrshort{h}&series\\
    \citet{narajewski_ensemble_2020}&15-19&\acrshort{mpf}&\acrshort{h}&series\\
    \citet{hirsch_simulation-based_2024}&16-20&\acrshort{mpf}&\acrshort{h}&series\\
    \citet{hirsch_multivariate_2024}&20-22&\acrshort{mpf}&\acrshort{h}&series\\
    \citet{scholz_towards_2021}&18&\acrshort{mpf}&\acrshort{h}&series\\
    \citet{hornek_comparative_2024}&21-22&\acrshort{mpf}&\acrshort{h}&series\\
    \midrule
    \textbf{Our study}&\textbf{24-25}&\textbf{\acrshort{mpf}}&\textbf{\acrshort{h}}&\textbf{series} \\
    \bottomrule
    \end{tabular}
    \begin{tablenotes}
            \footnotesize
            \item[*] Difference to \gls{da} price.
        \end{tablenotes}
    \end{threeparttable}
\end{table}

Table~\ref{tab:epf_literature_overview} provides an overview of the literature which analyzes \gls{epf} in the German \gls{cid} market.
For each study, we document the time period of the data used, including both the fitting and testing intervals of the models.
Additionally, we detail approaches to \gls{epf}, which include \acrlong{spf} (\acrshort{spf}), \acrlong{tf} (\acrshort{tf}), or \acrlong{mpf} (\acrshort{mpf}).
We also specify the forecasted \gls{cid} products, be they \acrlong{qh} (\acrshort{qh}) or \acrlong{h} (\acrshort{h}) products, and we identify the relevant forecasting targets.
In \acrlong{spf} studies, researchers typically employ the \gls{vwap} of all transactions, namely, the \gls{idfull}, or the \gls{vwap} of transactions during the previous three hours of trading, namely the \gls{id3}~\cite{epex_spot_se_description_2023}.
Conversely, \acrlong{mpf} and \acrlong{tf} require there to be multiple targets for each product, specifically with them using price \textit{series} that consist of several sequential \glspl{vwap}.

Predominantly, the studies analyze activity from 2015 to the start of the COVID-19 crisis, that is, up to 2019.
The majority of the research focuses on questions around \acrlong{spf}, with the \gls{id3} price being scrutinized in particular.
More research has been devoted to exploring \acrlong{h} products than to those using \acrlong{qh} data.
This could be due to \acrlong{h} products having more market liquidity.
Research on \gls{mpf} has recently gained in popularity, due to the growing significance of algorithmic trading~\cite{kuppelwieser_intraday_2023}.
\Gls{mpf} is particularly useful for algorithmic trading systems, as it continuously updates price forecasts throughout the \gls{cid} trading session by incorporating the latest data, including recent price changes.
This real-time forecasting can serve as data inputs for trading algorithms, thus helping to optimize both the timing and pricing of orders.

In our study, we examine the most recent period, including data from 2025.
We adopt the \gls{mpf} approach because it more accurately represents \gls{cid} market structure. 
This renders the forecast relevant for practical applications, including for algorithmic trading.
Our study concentrates on hourly products, thus aligning with most of the literature.
We do this because these products have the highest levels of market liquidity.
This is beneficial for modeling because it features greater trading activity and thus more data points that are useful for our analysis.

Table~\ref{tab:epf_mpf_literature_overview} provides an overview of studies employing a \gls{mpf} approach.
\begin{table}[h!]
    \caption{\Gls{mpf} literature overview with forecasting task, model type, use of \glspl{neighbor}, and consideration of \glspl{lob}.}
    \label{tab:epf_mpf_literature_overview}
    \centering
    \begin{tabular}{lcccc}
    \toprule
    Study & Task & Type & \acrshortpl{neighbor} &\acrshort{lob} \\
    \midrule
    \citet{narajewski_ensemble_2020} & \acrshort{reg} & \acrshort{prob} & \xmark & \xmark \\
    \citet{hirsch_simulation-based_2024} & \acrshort{reg} & \acrshort{prob} & \xmark & \xmark \\
    \citet{hirsch_multivariate_2024} & \acrshort{reg} & \acrshort{prob} & \acrshort{h} & \xmark \\
    \citet{scholz_towards_2021} & \acrshort{reg} & \acrshort{det} & \xmark & \cmark \\
    \citet{hornek_comparative_2024} & \acrshort{reg} & \acrshort{det} & \xmark & \xmark \\
    \midrule
    \textbf{Our study} & \textbf{\acrshort{class}} & \textbf{\acrshort{prob}} & \textbf{\acrshort{h}, \acrshort{qh}} & \cmark \\
    \bottomrule
    \end{tabular}
    \captionsetup{justification=centering}
\end{table}
While prior work has predominantly adopted a \gls{reg} paradigm, to our knowledge our study is the first to apply a \gls{class} framework. This choice aligns more closely with the needs of algorithmic trading, where directional price signals are often sufficient for decision-making.
The reviewed literature further differentiates between \gls{prob} and \gls{det} forecasting approaches.
\Gls{det} methods yield a single-point estimate, disregarding the inherent uncertainty and variability of market prices.
In contrast, \gls{prob} approaches account for these uncertainties by estimating distributions or probabilities over possible outcomes.
Given the stochastic nature of the \gls{cid} market, we adopt a \gls{prob} approach to better capture its dynamics.
To date, only \citet{hirsch_multivariate_2024} have considered dependencies on neighboring products, and even then, solely on \acrlong{h} (\acrshort{h}) products.
We extend this line of inquiry by incorporating both \acrlong{h} (\acrshort{h}) and \acrlong{qh} (\acrshort{qh}) neighboring products, which offer additional potential for improving forecast accuracy.
Finally, we investigate the role of features derived from the \gls{lob}—an aspect previously explored only by \citet{scholz_towards_2021}, and not yet studied within the context of directional \gls{epf}.
\section{Research Process and modeling}\label{sec:modeling}
We conduct our research following a design science paradigm~\cite{peffers_design_2007}.
Specifically, we develop a directional price forecasting method for hourly products in the European \gls{cid} market.
The proposed method generates probabilistic signals indicating the direction of market movements, which can be leveraged for algorithmic trading.

Section~\ref{sec:methodology} outlines the overall methodological approach grounded in the design paradigm.
Section~\ref{sec:forecasting_setup} presents the forecasting setup for a single \gls{cid} product.
Section~\ref{sec:forecasting_problem_definition} formally defines the directional price forecasting problem.
Section~\ref{sec:neighboring_products} addresses the challenge of incorporating information from neighboring products.
Section~\ref{sec:features} details the forecasting features, including those derived from neighboring products.
Section~\ref{sec:classification_models} introduces the classification models we use.
Finally, Section~\ref{sec:evaluation_metrics} details our evaluation metrics.

\subsection{Methodology}\label{sec:methodology}
Our study adopts a design process paradigm based on~\cite{peffers_design_2007} to develop and evaluate our forecasting approach method to create directional \gls{epf} in the \gls{cid}.
The process is iterative and consists of six main steps.
\begin{enumerate}
    \item \textbf{Problem Identification and Motivation:}
    We define the research problem within the context of a collaborative project with an energy supplier, aimed at analyzing the \gls{cid} market in the German-Luxembourgish region. To establish the state of the art, we conduct a narrative literature review following the guidelines of \cite{GREEN2006101} and ~\cite{pare_synthesizing_2015}. We present our findings in Section~\ref{sec:related_works_and_contributions}. Our literature analysis reveals a clear research gap: the limited investigation of directional \gls{epf} in the \gls{cid} market and the underuse of informative features, particularly those derived from neighboring products and \glspl{lob}.
    
    \item \textbf{Definition of Objectives for a Solution}: Our primary objective is to propose a directional \gls{epf} method that delivers high forecasting accuracy across multiple evaluation metrics. The secondary objective is to examine whether incorporating neighboring products—both hourly and quarter-hourly—as well as features derived from \glspl{lob}, can improve forecasting performance, and to quantify the extent of this improvement.
    
    \item \textbf{Design and Development of the Artifact:} The artifact we propose is a novel directional \gls{epf} method that formulates directional forecasting as a classification task based on \gls{cid} transaction data. The approach includes the construction of both endogenous and exogenous features, explicitly incorporating information from neighboring hourly and quarter-hourly products, as well as features derived from the \gls{lob}. The method involves data preprocessing and transformation to enable the application of \gls{ml} classification algorithms, which output both class labels and associated probabilities (i.e., indicating a rising or falling market). The implementation was developed in Python using the scikit-learn library~\cite{pedregosa_scikit-learn_2011}.
    
    \item \textbf{Demonstration:} We demonstrate the applicability of the proposed method using real-world data from \citet{epex_spot_se_market_2024}.
    
    \item \textbf{Evaluation:} We assessed the performance of our method through \gls{cv}, using evaluation metrics such as accuracy and \gls{pnl}, as further discussed in Section~\ref{sec:evaluation_metrics}. To test for statistical significance, we applied the \gls{dm} test~\cite{diebold_comparing_1995}. The insights gained from these analyses informed subsequent refinement cycles of the artifact.
    
    \item \textbf{Communication}: We present our artifact and its underlying design in this manuscript.
\end{enumerate}

Overall, the design process involved multiple refinement cycles between the design and evaluation stages.
The entire iterative development spanned a period of two years, with each iteration revisiting key aspects such as feature selection, model architecture, and hyperparameter configurations.
This manuscript presents the final iteration of the artifact resulting from this process.

\subsection{Forecasting Setup}\label{sec:forecasting_setup}
\begin{figure}[h!]
    \centering
    \begin{tikzpicture}[scale=0.8]
    \draw[thick, ->] (0.5,0) -- (9.5,0) node[anchor=west] {Time};

    \draw[thick] (7.5,0.1) -- (7.5,-0.1) node[below, yshift=-2pt] {$s$}; 
    \draw[thick] (9,0.1) -- (9,-0.1) node[below] {$s\!+\!\ell$};

    \draw[thick] (1,0.1) -- (1,-0.1) node[below] {$s\!-\!3h$};
    \draw[thick] (6,0.1) -- (6,-0.1) node[below] {$s\!-\!30min$};

    \draw[thick, decorate, decoration={brace, amplitude=10pt}] (1,0.3) -- (6,0.3)
        node [midway, above, yshift=10pt] {Forecasting Period};

    \draw[thick, decorate, decoration={brace, amplitude=10pt}] (7.5,0.3) -- (9,0.3)
        node [midway, above, yshift=10pt] {Delivery Period};
\end{tikzpicture}
    \caption{Forecasting overview for an exemplary product $(s,\ell)$.}
    \label{fig:forecasting_overview}
\end{figure}
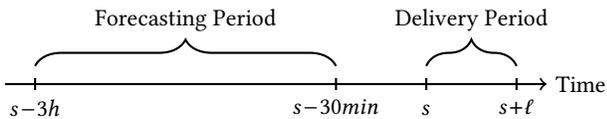

In the following, we detail our forecasting setup.  
We consider an arbitrary power product in the \gls{cid} market as a reference point for the subsequent explanations.  
In the \gls{cid} market, a product refers to a specified quantity of power scheduled for delivery over a defined time interval. 
We associate each product with two distinct periods: the forecasting period and the delivery period.  
The forecasting period is the time during which the method predicts the price of the product scheduled for delivery in the delivery period.  
We denote \gls{cid} products based on their delivery periods using tuples $(s, \ell)$, where $s$ indicates the start of delivery and $\ell$ specifies the delivery duration. Figure~\ref{fig:forecasting_overview} depicts these two periods.

In our context, the delivery product start ($s$) corresponds to the beginning of any full hour, and the product delivery length ($\ell$) spans one hour, aligning with our focus on forecasting hourly product prices.
Note that the forecasting process can easily be generalized to \acrlong{qh} products where ($s$) would correspond to the beginning of any quarter hour, and the product delivery length ($\ell$) would span 15 minutes.
However, forecasting prices for \acrlong{qh} products is beyond the scope of this study.
The forecasting period coincides with the \gls{id3} period.
The \gls{id3} is a price index widely used in industry for the \gls{cid} market which covers trades from three hours to 30 minutes before delivery~\cite{epex_spot_se_description_2023}.

\subsubsection{\Acrlong{rw}}
\begin{figure}[h!]
    \centering
    \begin{tikzpicture}
    \def\xscale{0.8}
    \def\yscale{0.7}

    \draw[->, thick] ({0.5*\xscale},{0*\yscale}) -- ({7.5*\xscale},{0*\yscale}) node[anchor=west] {Time};
    \foreach \y in {-2, -3, -4} {
        \draw[->, thick] ({0.5*\xscale},{\y*\yscale}) -- ({9.5*\xscale},{\y*\yscale});
    }

    \foreach \y in {0, -2, -3, -4} {
        \ifnum\y=0 
            \def\xmax{7}
        \else 
            \def\xmax{9}
        \fi;
        \foreach \x in {1,...,\xmax} {
            \draw[thick] ({\x*\xscale},{(\y+0.1)*\yscale}) -- ({\x*\xscale},{(\y-0.1)*\yscale});
        }
    }

    \foreach \y/\shift in {0/0, -2/1, -3/2, -4/3} {
        \fill[rounded corners, fill=red!30, fill opacity=0.5]
            ({(1+\shift)*\xscale},{(\y+0.2)*\yscale})
            rectangle
            ({(6+\shift)*\xscale},{(\y-0.2)*\yscale});
        \fill[red] ({(1+\shift)*\xscale},{\y*\yscale}) circle ({0.1*\xscale});
    }

    \draw[thick, decorate, decoration={brace, amplitude=\yscale*10pt}]
        ({2*\xscale},{-4.2*\yscale}) -- ({1*\xscale},{-4.2*\yscale})
        node [midway, below, yshift=-10pt*\yscale] {$1min$};

    \node[left] at ({0.5*\xscale},{0.0*\yscale}) {Step 1};
    \node[left] at ({0.5*\xscale},{-2.0*\yscale}) {Step 143};
    \node[left] at ({0.5*\xscale},{-3.0*\yscale}) {Step 144};
    \node[left] at ({0.5*\xscale},{-4.0*\yscale}) {Step 145};

    \node[above] at ({1*\xscale},{0.2*\yscale}) {$s\!-\!3h$};
    \node[above] at ({2*\xscale},{-1.8*\yscale}) {$s\!-\!37min$};
    \node[above] at ({3*\xscale},{-2.8*\yscale}) {$s\!-\!36min$};
    \node[above] at ({4*\xscale},{-3.8*\yscale}) {$s\!-\!35min$};
    \node[above] at ({9*\xscale},{-3.8*\yscale}) {$s\!-\!30min$};

    \foreach \y in {-0.75, -1.0, -1.25} {
        \filldraw ({3.5*\xscale},{\y*\yscale}) circle (1pt);
    }

    \matrix [draw, below of=axis, node distance=0.5cm, matrix of nodes, column sep=0.5cm] at (current bounding box.south) {
        \node[draw, fill=red, circle, label=right:Forecasting time] {}; &
        \node[draw, fill=red!30, label=right:Forecasted period] {}; \\
    };
\end{tikzpicture}
    \caption{\acrshort{rw} forecasting overview.}
    \label{fig:rw_forecasting_overview}
\end{figure}
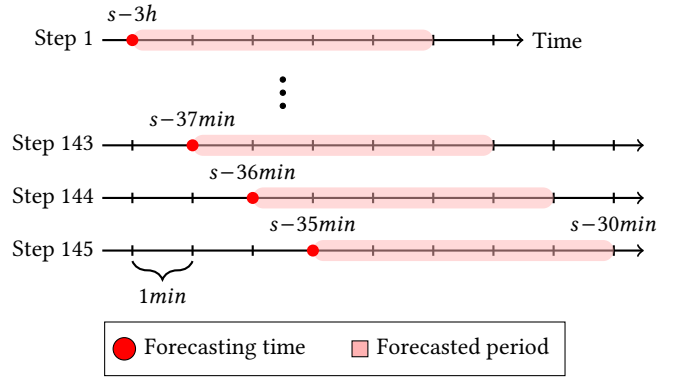

During the forecasting period, we apply a \gls{rw} forecasting approach illustrated in Figure~\ref{fig:rw_forecasting_overview}.
Within the \gls{rw} forecasting setup, there are two important moments in time: the forecasting time, which is the moment the forecast is made, and the forecasted period, which is the period for which a forecast is made.

We make the first forecast (Step 1) three hours before the delivery start ($s-3h$), and the final forecast (Step 145) 35 minutes before the delivery start ($s$), thus amounting to a total of 145 forecasts for each product.
Note that we make the last forecast five minutes before the end of the forecast period ($s-35min$), because extending the \gls{rw} further would result in the five-minute interval encompassing moments in time beyond the end of the forecast period, which concludes 30 minutes before the delivery start ($s-30min$).

For example, if the delivery period is 5:00 AM to 6:00 AM, we would start to run the method at 2:00 AM, forecasting the direction of the prices of the within the five-minute interval from 2:00 AM to 2:05 AM.
We would then run it again at 2:01 AM, forecasting the prices for the 5-minute interval from 2:01 AM to 2:06 AM, and so on, with the last forecast occurring at 5:25 AM for the period from 5:25 AM to 5:30 AM.

\subsubsection{\Acrlong{cv}}
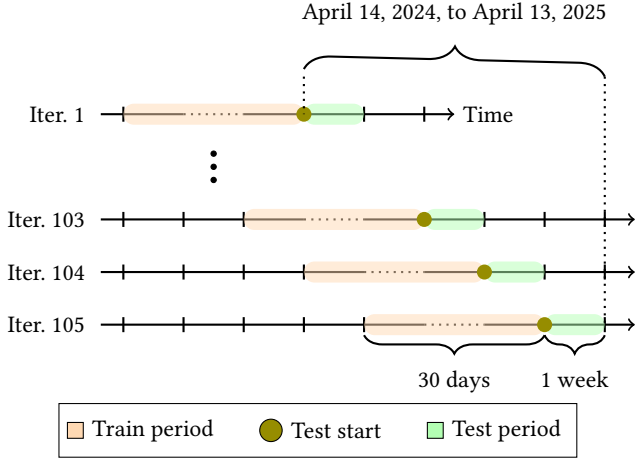
\begin{figure}[h!]
    \centering
    \begin{tikzpicture}
    \def\xscale{0.8}
    \def\yscale{0.7}

    \foreach \xmax/\y in {6.5/0, 9.5/-2, 9.5/-3, 9.5/-4} {
        \draw[-, thick] (0.5,\y*\yscale) -- (-\y*\xscale+2*\xscale,\y*\yscale);
        \draw[-, thick, dotted] (-\y*\xscale+2*\xscale,\y*\yscale) -- ((-\y*\xscale+3*\xscale,\y*\yscale);
        \draw[->, thick] (-\y*\xscale+3*\xscale,\y*\yscale) -- (\xmax*\xscale,\y*\yscale) \ifnum\y=0 node[anchor=west] {Time} \fi;
    }

    \foreach \xenda/\xendb/\y in {1/2/0, 3/5/-2, 4/5/-3, 5/5/-4} {
        \foreach \x in {1,...,\xenda} {
            \draw[thick] (\x*\xscale,\y*\yscale+0.1) -- (\x*\xscale,\y*\yscale-0.1);
        }
        \foreach \x in {\xenda,...,\xendb} {
            \draw[thick] (\x*\xscale+4*\xscale,\y*\yscale+0.1) -- (\x*\xscale+4*\xscale,\y*\yscale-0.1);
        }
    }

    \foreach \y/\shift in {0/3, -2/5, -3/6, -4/7} {
        \fill[rounded corners, fill=orange!30, fill opacity=0.5]
            (1*\xscale+\shift*\xscale,\y*\yscale+0.2*\yscale) rectangle (-2*\xscale+\shift*\xscale,\y*\yscale-0.2*\yscale);
        \fill[rounded corners, fill=green!30, fill opacity=0.5]
            (1*\xscale+\shift*\xscale,\y*\yscale+0.2*\yscale) rectangle (2*\xscale+\shift*\xscale,\y*\yscale-0.2*\yscale);
        \fill[olive] (1*\xscale+\shift*\xscale,\y*\yscale) circle (0.1);
    }

    \draw[thick, decorate, decoration={brace, amplitude=\yscale*10pt}]
        (9*\xscale,-4.2*\yscale) -- (8*\xscale,-4.2*\yscale)
        node [midway, below, yshift=-10pt] {1 week};
        
    \draw[thick, decorate, decoration={brace, amplitude=\yscale*10pt}]
        (8*\xscale,-4.2*\yscale) -- (5*\xscale,-4.2*\yscale)
        node [midway, below, yshift=-10pt] {30 days};

    \draw[-, thick, dotted] (9*\xscale,0.5) -- (9*\xscale,-4*\yscale);
    \draw[-, thick, dotted] (4*\xscale,0.5) -- (4*\xscale,0*\yscale);
    \draw[thick, decorate, decoration={brace, amplitude=10pt}]
        (4*\xscale,0.6) -- (9*\xscale,0.5)
        node [midway, below, yshift=30pt] {April 14, 2024, to April 13, 2025};

    \node[left] at (0.5*\xscale,0.0*\yscale) {Iter. 1};
    \node[left] at (0.5*\xscale,-2.0*\yscale) {Iter. 103};
    \node[left] at (0.5*\xscale,-3.0*\yscale) {Iter. 104};
    \node[left] at (0.5*\xscale,-4.0*\yscale) {Iter. 105};

    \filldraw (2.5*\xscale,-0.75*\yscale) circle (1pt);
    \filldraw (2.5*\xscale,-1.0*\yscale) circle (1pt);
    \filldraw (2.5*\xscale,-1.25*\yscale) circle (1pt);

    \matrix [draw, below of=axis, node distance=.4cm, matrix of nodes, column sep=0.5cm] at (current bounding box.south) {
        \node[draw, fill=orange!30, label=right:Train period] {}; &
        \node[draw, fill=olive, circle, label=right:Test start] {}; &
        \node[draw, fill=green!30, label=right:Test period] {}; \\
    };
\end{tikzpicture}
    \caption{\acrshort{cv} overview.}
    \label{fig:cv_overview}
\end{figure}

To account for the temporal changes in the market, we apply a \gls{cv}.
A \gls{cv}, particularly in the context of time-series analysis, involves periodically retraining the forecasting method to adapt to evolving trends within the time-series data.
In the context of the \gls{cid} market analyzing temporal trends it is important to ensure that methods are able to adapt to changing market dynamics.
Figure~\ref{fig:cv_overview} illustrates the \gls{cv} procedure we employ in this study, wherein we divide the one year test period (April 14, 2024, to April 13, 2025) into 52 consecutive weekly intervals.  

For every one-week test interval we retrain the forecasting model using data from the preceding 30 days.
We insert a one-day buffer between training and testing to prevent contamination of the training set with future data.

\subsection{Forecasting Problem Definition}\label{sec:forecasting_problem_definition}

In each \gls{rw}, we aim to generate a directional forecast that predicts whether the market will rise or fall over the next five minutes. To achieve this, we formulate the problem as a classification task based on numerical price data.

We construct the classification target by comparing two price levels: a reference price and a future price, both computed using \glspl{vwap}.
To determine the reference price, we average the \glspl{vwap} of the four most recent transactions occurring before the forecast time following the authors of~\cite{hornek_comparative_2024}.
To obtain the future price, we calculate the average \gls{vwap} of all transactions within the following five-minute interval.

We then define the market direction by comparing these two values: if the future price exceeds the reference price, the market is rising; otherwise, it is falling.
By applying this method to all windows in the \gls{rw}, we construct a dataset suitable for classification.

\subsection{Neighboring Product Integration}\label{sec:neighboring_products}

Our forecasting method uses neighboring products with overlapping trading periods, necessitating adjustments in the modeling strategy. 
Since the delivery periods of neighboring products differ from that of the forecasted product, trading in some neighboring products may cease during the forecasting period, rendering trade data for these products unavailable.

\begin{figure}[h!]
    \centering
    \begin{tikzpicture}
    \def\xscale{0.8}
    \def\yscale{0.7}

    \foreach \shift/\len/\pattern in {0/1/vertical lines,1/1/bricks,2/0.5/crosshatch} {
        \draw[dashed, pattern=\pattern, pattern color=black!30, draw opacity=0.5] (\xscale*1+\xscale*\shift,\yscale*0.2-\yscale*\shift) rectangle (\xscale*1+\xscale*\len+\xscale*\shift,-\yscale*4.2);
    }

    \draw[-, thick, dotted] (\xscale*4, 0) -- (\xscale*4, \yscale*-4.6) node [below, yshift=-5pt] {$s$};

    \draw[->, thick] (\xscale*0.5,\yscale*0) -- (\xscale*5.5,\yscale*0) node[anchor=west] {Time};
    \foreach \y in {-1, ..., -4} {
        \draw[->, thick] (\xscale*0.5,\yscale*\y) -- (\xscale*7.5,\yscale*\y);
    }

    \foreach \y/\name in {1/Product,0/{($s-2h$,$1h$)}, -1/{($s-1h$,$1h$)}, -2/{\bm{$(s,1h)$}},  -3/{($s+1h$,$1h$)}, -4/{($s+2h$,$1h$)}} 
        \node[left] at (\xscale*0.5,\yscale*\y) {\name};

    \foreach \y in {0, -1, ..., -4} {
        \ifnum\y=0 
            \def\xmax{5}
        \else 
            \def\xmax{7}
        \fi;
        \foreach \x in {1,...,\xmax} {
            \draw[thick] (\xscale*\x,\yscale*\y+0.1) -- (\xscale*\x,\yscale*\y-0.1);
        }
    }

    \foreach \y/\x in {0/2, -1/3, -2/4, -3/5, -4/6} {
        \fill[rounded corners, fill=blue!30, fill opacity=0.5] (\xscale*\x,\yscale*\y+0.2) rectangle (\xscale*\x+\xscale*1,\yscale*\y-0.2);
        \fill[blue] (\xscale*\x,\yscale*\y) circle (0.1);
    }

    \draw[thick, decorate, decoration={brace, amplitude=\yscale*10pt}] (\xscale*2,-\yscale*4-0.2) -- (\xscale*1,-\yscale*4-0.2)
    node [midway, below, yshift=-10pt] {$1h$};

    \matrix [draw, below of=axis, node distance=1cm, matrix of nodes, column sep=0.2cm] at (current bounding box.south) {
        \node[draw, fill=blue, circle, label=right:Delivery start] {}; &
        \node[draw, fill=blue!30, label=right:Delivery period] {}; \\
        \node [draw, pattern=vertical lines, pattern color=black!30, dashed, draw opacity=0.5, minimum width=0.4cm, minimum height=0.4cm, label=right:Period $3h\!\rightarrow\!2h$] {}; &
        \node [draw, pattern=bricks, pattern color=black!30, dashed, draw opacity=0.5, minimum width=0.4cm, minimum height=0.4cm, label=right:Period $2h\!\rightarrow\!1h$] {};&
        \node [draw, pattern=crosshatch, pattern color=black!30, dashed, draw opacity=0.5, minimum width=0.4cm, minimum height=0.4cm, label=right:Period $1h\!\rightarrow\!\frac{h}{2}$] {};\\
    };

\end{tikzpicture}
    \caption{Overview of parallel hourly products.}
    \label{fig:parallel_product_overview_hours}
\end{figure}
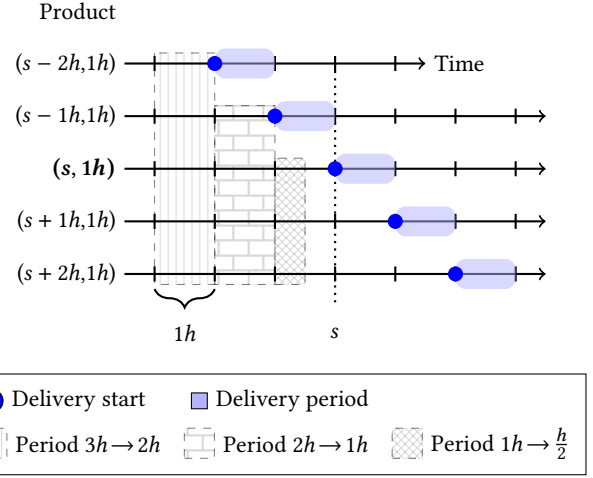

Figure~\ref{fig:parallel_product_overview_hours} illustrates the timelines of various hourly products that trade concurrently with the current hourly product $(s,1h)$—the product for which we make forecasts. 
In the figure, a dot marks the delivery start of each product, and the adjacent shaded area denotes the corresponding one-hour delivery period.

We consider a total of four neighboring hourly products: two with delivery starting before ($(s-2h,1h)$ and $(s-1h,1h)$) and two after ($(s+1h,1h)$ and $(s+2h,1h)$) the delivery of the current product. 
Products whose delivery starts within the trading period of the current product present modeling challenges, as no further price data becomes available after delivery begins.
This is particularly relevant, as price dynamics near delivery tend to be more informative due to increased market liquidity and a closer alignment of prices with delivery conditions.

To address this issue, we divide the forecasting period into three periods, each aligned with the delivery start times of specific hourly products:
\begin{itemize}[nosep]
    \item \textbf{Period $3h \rightarrow 2h$}: Includes $(s-2h,1h)$, $(s-1h,1h)$, $(s+1h,1h)$, and $(s+2h,1h)$.
    \item \textbf{Period $2h \rightarrow 1h$}: Includes $(s-1h,1h)$, $(s+1h,1h)$, and $(s+2h,1h)$.
    \item \textbf{Period $1h \rightarrow \tfrac{1}{2}h$}: Includes $(s+1h,1h)$ and $(s+2h,1h)$.
\end{itemize}

The naming of each period refers to the time remaining until the delivery of the current product.
For example, the period $3h \rightarrow 2h$ spans the interval between three and two hours prior to delivery start of the current product $(s,1h)$.

In addition to hourly products, we also incorporate quarter-hourly products.
For consistency, we retain the same period segmentation as introduced for the neighboring hourly products.
In each period, we include the quarter-hourly products corresponding to the hourly products considered in that period.

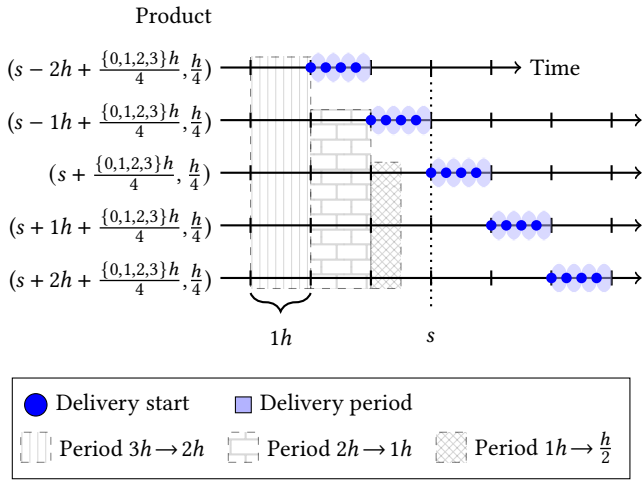
\begin{figure}[h!]
    \centering
    \begin{tikzpicture}
    \def\xscale{0.8}
    \def\yscale{0.7}

    \foreach \shift/\len/\pattern in {0/1/vertical lines,1/1/bricks,2/0.5/crosshatch} {
        \draw[dashed, pattern=\pattern, pattern color=black!30, draw opacity=0.5] (\xscale*1+\xscale*\shift,\yscale*0.2-\yscale*\shift) rectangle (\xscale*1+\xscale*\len+\xscale*\shift,-\yscale*4.2);
    }

    \draw[-, thick, dotted] (\xscale*4, 0) -- (\xscale*4, \yscale*-4.6) node [below, yshift=-5pt] {$s$};

    \draw[->, thick] (\xscale*0.5,\yscale*0) -- (\xscale*5.5,\yscale*0) node[anchor=west] {Time};
    \foreach \y in {-1, ..., -4} {
        \draw[->, thick] (\xscale*0.5,\yscale*\y) -- (\xscale*7.5,\yscale*\y);
    }

    \foreach \y/\name in {1/Product,0/{($s-2h+\frac{\{0,1,2,3\}h}{4}$,$\frac{h}{4}$)}, -1/{($s-1h+\frac{\{0,1,2,3\}h}{4}$,$\frac{h}{4}$)}, -2/{$(s+\frac{\{0,1,2,3\}h}{4},\frac{h}{4})$},  -3/{($s+1h+\frac{\{0,1,2,3\}h}{4}$,$\frac{h}{4}$)}, -4/{($s+2h+\frac{\{0,1,2,3\}h}{4}$,$\frac{h}{4}$)}} 
        \node[left] at (\xscale*0.5,\yscale*\y) {\name};

    \foreach \y in {0, -1, ..., -4} {
        \ifnum\y=0 
            \def\xmax{5}
        \else 
            \def\xmax{7}
        \fi;
        \foreach \x in {1,...,\xmax} {
            \draw[thick] (\xscale*\x,\yscale*\y+0.1) -- (\xscale*\x,\yscale*\y-0.1);
        }
    }

    \foreach \y/\startx in {0/2, -1/3, -2/4, -3/5, -4/6} {
        \foreach \i in {0,1,2,3} {
            \pgfmathsetmacro{\xstart}{\startx + 0.25*\i}
            \pgfmathsetmacro{\xend}{\xstart + 0.25}
            \fill[rounded corners, fill=blue!30, fill opacity=0.5] 
                (\xscale*\xstart,\yscale*\y+0.2) rectangle (\xscale*\xend,\yscale*\y-0.2);
            \fill[blue] (\xscale*\xstart,\yscale*\y) circle (0.07);
        }
    }


    \draw[thick, decorate, decoration={brace, amplitude=\yscale*10pt}] (\xscale*2,-\yscale*4-0.2) -- (\xscale*1,-\yscale*4-0.2)
    node [midway, below, yshift=-10pt] {$1h$};

    \matrix [draw, below of=axis, node distance=1cm, matrix of nodes, column sep=0.2cm] at (current bounding box.south) {
        \node[draw, fill=blue, circle, label=right:Delivery start] {}; &
        \node[draw, fill=blue!30, label=right:Delivery period] {}; \\
        \node [draw, pattern=vertical lines, pattern color=black!30, dashed, draw opacity=0.5, minimum width=0.4cm, minimum height=0.4cm, label=right:Period $3h\!\rightarrow\!2h$] {}; &
        \node [draw, pattern=bricks, pattern color=black!30, dashed, draw opacity=0.5, minimum width=0.4cm, minimum height=0.4cm, label=right:Period $2h\!\rightarrow\!1h$] {};&
        \node [draw, pattern=crosshatch, pattern color=black!30, dashed, draw opacity=0.5, minimum width=0.4cm, minimum height=0.4cm, label=right:Period $1h\!\rightarrow\!\frac{h}{2}$] {};\\
    };

\end{tikzpicture}
    \caption{Overview of parallel quarter-hourly products.}
    \label{fig:parallel_product_overview_quarterhours}
\end{figure}

We consider quarter-hourly products for both neighboring and current hourly products.
Figure~\ref{fig:parallel_product_overview_quarterhours} illustrates the quarter-hourly products we consider.
For example, for the neighboring hourly product $(s-1h,1h)$, we include the corresponding quarter-hourly products: $(s-1h,\tfrac{h}{4})$, $(s-1h+\tfrac{h}{4},\tfrac{h}{4})$, $(s-1h+\tfrac{2h}{4},\tfrac{h}{4})$, and $(s-1h+\tfrac{3h}{4},\tfrac{h}{4})$.
Similar to the hourly product $(s-1h,1h)$, we consider these quarter-hourly products only during the periods $3h\rightarrow 2h$ and $2h \rightarrow 1h$.

As a consequence of the period-wise segmentation, we train separate \gls{ml} models for each period, as the set of input features evolves throughout periods.

\subsection{Features}\label{sec:features}
In this section we introduce the features we use in the study: \gls{cid} prices derived from transactions and \gls{lob}, the imbalance, and fundamental variables, including load and \gls{vre} generation.

\subsubsection{\acrshort{cid} Prices}\label{sec:ftr_cid_prices}
\gls{cid} products trade continuously, with trades occurring at random intervals.
Consequently, one must discretize the trade data to use them as features in any forecasting method.
Our discretization strategy relies on two components: the most recent price information and the historical price trajectory.

For the most recent price information, we use the \gls{vwap} calculated from the last four trades~\cite{hornek_comparative_2024}.
To capture the historical price development, we construct a price vector using multiple lagged \glspl{vwap}, each representing trades aggregated over one-minute intervals.
With a maximum lag of $h_{\max}\in\mathbb{N}$, this results in a vector comprising $h_{\max}$ \glspl{vwap}, containing the price history for the preceding $h_{\max}$ minutes.
Including lagged terms in the method improves the forecast by enabling the method to learn temporal dependencies from past data.

For the current product, we include both the most recent price information and the historical price vector, resulting in \(h_{\max}+1\) price features.
For neighboring products, we use only the historical price vector, yielding \(h_{\max}\) features for each neighboring product.
We omit the most recent price information for neighboring products because empirical experiments indicate that no significant improvement in accuracy accrues when these prices were included.
We compute the historical price vector for neighboring products until the delivery start (see Figure~\ref{fig:parallel_product_overview_hours}).
During the \gls{sdat} period (30 to 5 minutes before delivery, see Figure~\ref{fig:trading_timeline_h}), we average transactions across all German-Luxembourgish delivery areas to create an aggregate price feature.
This averaging helps combine the transactions in four order books into one price feature for the entire German-Luxembourgish delivery area.
We then forward-fill price features during the remaining five minutes before delivery.

We determine \( h_{\max} = 10 \) lags for the price history of any product based on experimental results.
For additional details regarding the computation of these price features, refer to Appendix~\ref{app:cid_price_ftr}.
We source the \gls{cid} transactions from the \gls{epex} SFTP server~\cite{epex_spot_se_market_2024}.


\subsubsection{\Acrlong{lob}}
In addition to historical trades, we also use the current \gls{lob} as a source of price-related features.
The \gls{lob} contains remaining limit orders, which are orders to buy (bid) or sell (ask) a certain product at specific prices or better (same price or lower for a bid, or same price or higher for an ask).
We compute features as \glspl{vwap} at different depths of the \gls{lob}.
More precisely we look at different numbers of top orders, focusing on those with the highest bid prices and the lowest ask prices, which are the most likely to be executed.

We apply two selection methods: (1) by number of orders and (2) by cumulative volume.
We use the top 1, 5, and 10 orders, or the top 1, 5, and 10~MW, respectively, as we did not have sufficient time to explore additional parametrizations.
For each side of the order book (bid or ask), we obtain three features for each selection method.

\subsubsection{Price Feature Normalization}\label{sec:ftr_price_norm}
The existing literature on \gls{epf} indicates that normalization followed by variance stabilizing transformations can improve forecasting accuracy, particularly in the context of \gls{da} price forecasting~\cite{uniejewski_variance_2018}.
However, there is a notable gap in the literature concerning variance stabilization for short-term \gls{cid} \gls{epf}.
Following~\cite{hornek_comparative_2024}, we applied a \gls{rw} normalization approach to a set of simple price forecasts, allowing these forecasts to remain comparable over time despite changing market environments.
We consider our preprocessing of price data essential for modeling, given the following two reasons:

First, general price levels, often measured using the \gls{id3} in the \gls{cid} market, exhibit significant trends over time.
Consequently, methods trained on untransformed data may exhibit a bias towards periods experiencing higher prices and under-perform during periods of lower prices.
This is due to differences in error magnitudes.
Second, some \gls{ml} models, such as linear models, perform better with normalized inputs~\cite{SINGH2020105524}.

A common approach for price data transformation in the literature for \gls{cid} \gls{epf} for detrending time-series is differentiation~\cite{narajewski_ensemble_2020,hirsch_simulation-based_2024,hirsch_multivariate_2024}.
However, differentiation does not address changing price volatility, yielding price data where variance changes over time.
This aspect is particularly important in the context of \gls{cid} trading because volatility changes both within trading sessions~\cite{narajewski_ensemble_2020} and across products~\cite{kulakov_impact_2021}.
To address these concerns, we adopt a \gls{rw} price normalization procedure for all price features, including both \gls{cid} and \gls{da} prices.
This approach expands upon the existing method of price normalization introduced by~\cite{hornek_comparative_2024}.

Our price data transformation approach consists of two steps and repeats at any distance to delivery, i.e., at every step of the \gls{rw} forecasts (every minute in this study's setup).
We normalize prices into standard scores, using \gls{vwap} and \gls{vwsd} (see Appendix~\ref{app:cid_notation}) as the parameters for mean and standard deviation, respectively~\cite{hornek_comparative_2024}.
To compute the \gls{vwap} and \gls{vwsd} we use all transactions known up to the forecasting time (see Appendix~\ref{app:cid_notation}).

\subsubsection{Load and Production Forecasts}
In addition to price-related features, we study fundamental features, focusing on forecasted load and \gls{vre} generation.
\Gls{vre} generation is mainly categorized into solar, wind onshore, and wind offshore~\cite{irena2024taxonomy}.
We source data for the German-Luxembourgish bidding zone from the \gls{entsoe} transparency platform~\cite{entso-e_entso-e_2025}.
The platform publishes two forecasts: a day-ahead forecast and an intraday forecast, with the latter updating the former.
For load, however, the platform provides only a day-ahead forecast and no intraday update.

\subsubsection{Imbalance}\label{sec:ftr_nrv}
In addition to price features, our analysis includes the current imbalance status of the power system.
We use the \gls{nrv}-saldo, also known as the "grid control cooperation balance" in English~\cite{regelleistung_gcc}.
The \gls{nrv}-saldo indicates the difference between consumption and generation, across the four German \glspl{tso}, whose absolute value the \glspl{tso} aim to minimize through the activation of balancing services~\cite{ku_leuven_energy_institute_ei_2015}.

The \glspl{tso} release the \gls{nrv}-Saldo data with a \acrlong{qh} resolution within a maximum delay of 30 minutes~\cite{50hertz_transmission_gmbh_nrv-saldo_2024}.
For example, the \gls{nrv} -Saldo for the period from 15:45 to 16:00 is made available no later than 16:15.
Our method incorporates the NRV-Saldo by using the most recent value from the latest available 15-minute period.

\subsection{Classification Models}\label{sec:classification_models}
In our study, we propose a method for \textit{directional} \gls{epf}, aiming to forecast the market direction.
Within the context of \gls{ml}, this task corresponds to a binary classification problem: the market either rises or falls.
To generate directional forecasts from the features, we train and test different \gls{ml} classification models.
We select models that not only predict the market direction but also provide an associated probability for the forecasted class.
Such probabilistic information can enhance risk management in downstream applications and support model evaluation by distinguishing forecasts based on the model's certainty.

Specifically, we consider and compare two \gls{ml} classification models: a logistic regression model (implemented via scikit-learn~\cite{pedregosa_scikit-learn_2011}) and a gradient boosting model (LightGBM~\cite{ke_lightgbm_2017}).
We choose logistic regression for its simplicity as a linear model and computational efficiency, and LightGBM to capture non-linear relationships while maintaining practical computational requirements.

However, particular characteristics of our dataset require adaptations to these models.
In particular, our price features exhibit substantial multicollinearity, primarily caused by autoregressive lags and correlated movements across products.
If left unaddressed, multicollinearity can impair model performance.
To mitigate this issue, we either adapt the regression method or transform the features.

For logistic regression, we apply \textit{l2} regularization, which distributes weights across correlated variables and reduces the adverse effects of multicollinearity.
Since gradient boosting does not inherently address multicollinearity, we first transform the feature set using the \glspl{pls} method\footnote{Although \glspl{pca} are often used to reduce dimensionality and produce orthogonal features, they fail to prioritize variables based on temporal relevance.
In our setting, more recent data hold greater predictive importance than older observations.
We therefore use \gls{pls}, a supervised alternative to \gls{pca}, which emphasizes features that are most predictive of the target~\cite{hastie2009elements}.
To further improve the decomposition quality, we fit the \gls{pls} model using the \gls{vwap} over the next five minutes rather than the classification target.}—retaining all components—before fitting the model~\cite{hastie2009elements}.

\subsection{Evaluation metrics}\label{sec:evaluation_metrics}
To evaluate our forecasting method, we apply two metrics: accuracy and \gls{pnl}.
Accuracy is widely used in \gls{ml}~\cite{hastie2009elements} and captures the proportion of test samples for which the model correctly predicts market direction.
The \gls{pnl} reflects the forecast's market value if executed on the market.

We simulate a trading strategy where we open a position at the reference price—defined as the \gls{vwap} of the last four trades—and close it at the future price—the \gls{vwap} of trades within the next five minutes.
These intervals align with the classification problem's definition (see Section~\ref{sec:forecasting_problem_definition}). 
If the model predicts a market decline, we simulate selling at the reference price and buying at the future price.
For a predicted increase, we reverse this logic.
Correct forecasts result in profits (positive differences), while incorrect ones yield losses (negative differences).

Although we use a past reference price that is not tradeable in reality, this setup allows us to evaluate forecasts not only by their directional correctness but also by the magnitude of price changes.
Consider the following situation.
We produce three forecasts, being (1) a correct forecast with a \euro{1} price change, (2) a correct forecast with a \euro{2} change, and (3) an incorrect forecast with a \euro{10} change.
The accuracy metric treats each prediction equally, resulting in a 66\% accuracy.
However, the corresponding \gls{pnl} equals a loss of \euro{7}, emphasizing the importance of evaluating forecasts using the \gls{pnl} as well.
\section{Results and Discussion}\label{sec:results_discussion}
We demonstrate and evaluate our method using real data from the \gls{epex} for the German \gls{cid} market, with the test period spanning from April 14, 2024, to April 13, 2025.
Section~\ref{sec:eval_feature_sets} analyses various feature sets.
Section~\ref{sec:eval_models} compares and discusses different forecasting models.
Section~\ref{sec:eval_temp} analyzes the temporal variation in performance across the test period.
Section~\ref{sec:eval_summary} summarizes our findings.
Finally, Section~\ref{sec:limitations} outlines the limitations of the current study and highlights directions for future research.

\subsection{Feature sets}\label{sec:eval_feature_sets}
Table~\ref{tab:acc_feature_inclusion} reports gradient boosting model accuracy across three forecasting periods—$3h\rightarrow2h$, $2h\rightarrow1h$, and $1h\rightarrow\tfrac{h}{2}$—as defined in Section~\ref{sec:neighboring_products}.
We measure accuracy as the proportion of test samples for which the model correctly predicts the market direction.
We omit results for the logistic regression model due to overall lower performance and to preserve brevity.
The “Current” feature set, comprising only historical prices of the target product, serves as the baseline.
All other sets incorporate additional exogenous information, including fundamentals (load and production forecasts), prices of neighboring products (at hourly or quarter-hourly granularity), imbalances, and order book data.
Entries we marked as “N/A” indicate feature unavailability due to the corresponding products not being traded during the relevant periods.
The "Selected" feature set combines various exogenous variables—specifically, those that improve forecasting accuracy when added to the "Current" feature set.
In essence, we merge several promising feature sets in an attempt to further enhance forecasting accuracy.
To evaluate improvements in forecasting accuracy, we use the \gls{dm} test results (see Appendix~\ref{app:dm_testing_cid_epf}) to check for statistically significant gains and base the inclusion or exclusion of feature sets on the test results.
During the $3h \rightarrow 2h$ period, the "Fundamentals", "H Neighbors ($+1h$)", "H Neighbors ($+2h$)", "QH Neighbors ($+2h$)" and "Imbalance" sets are excluded.
In the subsequent $2h \rightarrow 1h$ period, the same sets except "H Neighbors ($+2h$)" are excluded again.
In the final $1h \rightarrow \tfrac{h}{2}$ period, only "QH (Current)", "QH Neighbors ($+2h$)", "LOB (top rows)", and "LOB (top MW)" are retained.
For detailed \gls{dm} test results, refer to Appendix~\ref{app:dm_test_results}.
\begin{table}[htb]
    \centering
    \caption{Gradient boosting model accuracy in \% under varying feature set inclusions.}
    \label{tab:acc_feature_inclusion}
    \begin{tabular}{lrrr}
\toprule
\multirow{2}{*}{Feature Set} & \multicolumn{3}{c}{Period} \\
& $3h\!\rightarrow\!2h$ & $2h\!\rightarrow\!1h$ & $1h\!\rightarrow\!\tfrac{h}{2}$ \\
\midrule
Current & 51.73 & 51.26 & 53.00 \\
Current + Fundamentals & 51.64 & 51.30 & 53.07 \\
Current + H Neighbors ($-1h$) & 53.31 & 52.45 & N/A \\
Current + H Neighbors ($-2h$) & 53.20 & N/A & N/A \\
Current + H Neighbors ($+1h$) & 51.61 & 50.81 & 52.70 \\
Current + H Neighbors ($+2h$) & 51.26 & 50.69 & 52.72 \\
Current + QH (Current) & 53.57 & 52.36 & 53.59 \\
Current + QH Neighbors ($-1h$) & 54.25 & 53.24 & N/A \\
Current + QH Neighbors ($-2h$) & 54.08 & N/A & N/A \\
Current + QH Neighbors ($+1h$) & 52.37 & 51.70 & 53.18 \\
Current + QH Neighbors ($+2h$) & 51.06 & 51.56 & 53.29 \\
Current + Imbalance & 51.64 & 51.25 & 52.96 \\
Current + LOB (top rows) & 58.04 & \textbf{55.93} & 56.87 \\
Current + LOB (top MW) & \textbf{58.22} & 55.91 & \textbf{57.30} \\
Current + Selected & 54.55 & 55.18 & 56.36 \\
\bottomrule
\end{tabular}

\end{table}

Features based on hourly and quarter-hour neighbors improve performance most in periods longer lead times, with quarter-hourly neighbors consistently yielding higher accuracy gains than hourly neighbors.
However, when these neighbors start delivery after the current product (e.g., $+1h$ or $+2h$), the improvement is small or even leads to a decrease in accuracy.
The addition of fundamental data results in decreases accuracy.
Inclusion of \gls{lob} features leads to the highest gains in accuracy, with "top MW" outperforming "top rows" in all periods except in period $2h\rightarrow1h$.
Lastly, the "Selected" feature set improves accuracy over the reference across all periods, however its accuracy does not exceed that of the models using \gls{lob} features.

The results indicate that short-term market forecasting is predominantly price-driven, as supported by notable improvements in accuracy when including \gls{lob} features.
Among neighboring product features, those products whose delivery begins within the current trading period the improvement in forecasting accuracy is higher than for products whose delivery begins after the delivery of the current product.
This difference suggests that neighboring products with shorter lead times—typically more liquid than the current product—offer more relevant information than those with longer lead times.
In contrast, fundamental and imbalance features do not contribute to predictive performance.
This limited impact is likely due to their outdated nature: imbalance data can lag by up to 30 minutes, and fundamental inputs such as load and \gls{vre} forecasts remain static in the current setup, reducing their value in capturing short-term market dynamics.

\subsection{Forecasting models}\label{sec:eval_models}
Although we evaluate both considered forecasting models—a logistic regression model and a gradient boosting model—under various feature sets, we report only the results for the best-performing feature sets from Section~\ref{sec:eval_feature_sets}.
These are as follows:
\begin{itemize}
    \item $3h\rightarrow2h$: Current + LOB (top MW)
    \item $2h\rightarrow1h$: Current + LOB (top rows)
    \item $1h\rightarrow\tfrac{h}{2}$: Current + LOB (top MW)
\end{itemize}

For a more fine-grained analysis, we incorporate the probabilistic information provided by the forecasting models into our evaluation.
Each forecast consists of two components: a predicted direction and an associated probability.
The probability is always at least 50\% for the respective direction, as otherwise the forecasted direction would be reversed.
We refer to this probability value as the signal strength.
In our analysis, we sort all test samples by decreasing signal strength and evaluate performance across increasing percentiles—starting with the top 1\%, then the top 2\%, and continuing up to the full test set.

\begin{figure}[htbp]
  \centering
  \begin{subfigure}[b]{\columnwidth}
    \includegraphics[width=\linewidth]{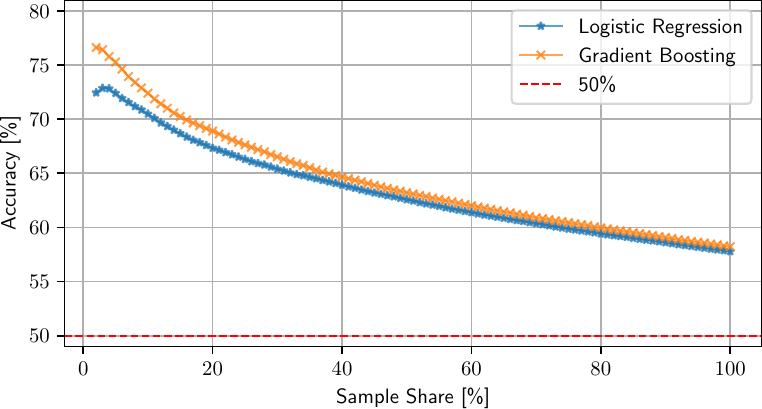}
    \caption{Period $3h\rightarrow2h$}
    \label{fig:acc_period_a}
  \end{subfigure}
  \hfill
  \vspace{0.0cm}
  \begin{subfigure}[b]{\columnwidth}
    \includegraphics[width=\linewidth]{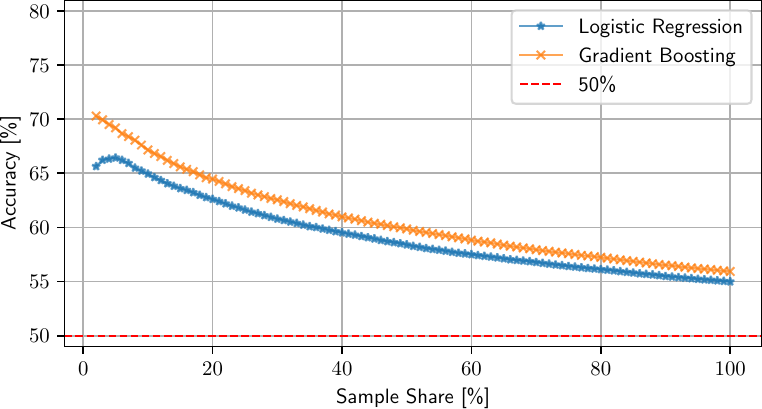}
    \caption{Period $2h\rightarrow1h$}
    \label{fig:acc_period_b}
  \end{subfigure}
  \hfill
  \vspace{0.0cm}
  \begin{subfigure}[b]{\columnwidth}
    \includegraphics[width=\linewidth]{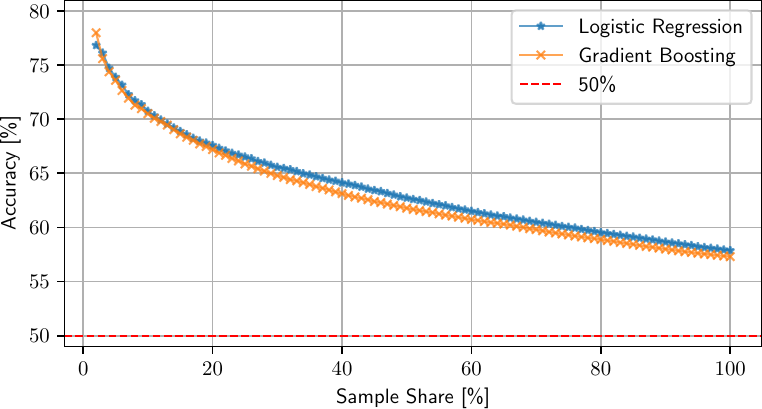}
    \caption{Period $1h\rightarrow\tfrac{h}{2}$}
    \label{fig:acc_period_c}
  \end{subfigure}
  \caption{Test accuracy across periods for varying sample shares with the highest signal strengths using the best-performing feature set across the test interval.}
  \label{fig:acc_period_a_b_c}
\end{figure}

\begin{figure}[htbp]
  \centering
  \begin{subfigure}[b]{\columnwidth}
    \includegraphics[width=\linewidth]{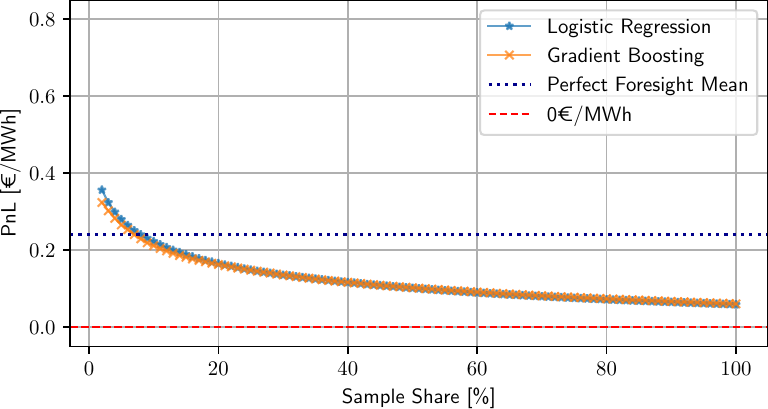}
    \caption{Period $3h\rightarrow2h$}
    \label{fig:pnl_period_a}
  \end{subfigure}
  \hfill
  \vspace{0.0cm}
  \begin{subfigure}[b]{\columnwidth}
    \includegraphics[width=\linewidth]{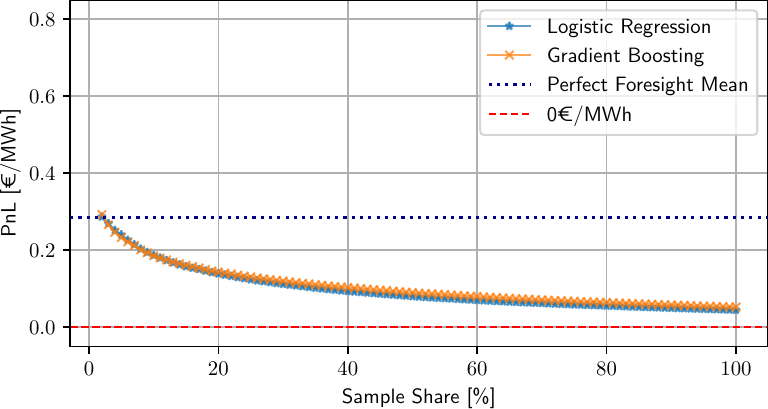}
    \caption{Period $2h\rightarrow1h$}
    \label{fig:pnl_period_b}
  \end{subfigure}
  \hfill
  \vspace{0.0cm}
  \begin{subfigure}[b]{\columnwidth}
    \includegraphics[width=\linewidth]{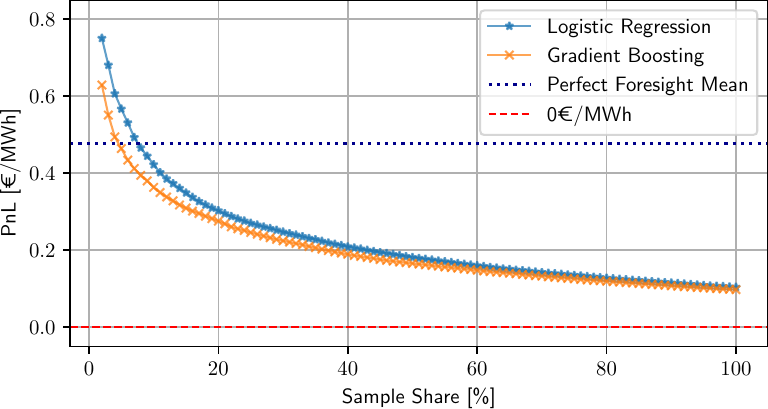}
    \caption{Period $1h\rightarrow\tfrac{h}{2}$}
    \label{fig:pnl_period_c}
  \end{subfigure}
  \caption{Mean \gls{pnl} across periods for varying sample shares with the highest signal strengths using the best-performing feature set across the test interval.}
  \label{fig:pnl_period_a_b_c}
\end{figure}

Figures~\ref{fig:acc_period_a_b_c} and \ref{fig:pnl_period_a_b_c} report the results for accuracy and \gls{pnl} (see Section~\ref{sec:evaluation_metrics}), respectively.
In Figure~\ref{fig:pnl_period_a_b_c}, we also include the mean absolute price difference across all samples as a benchmark.
This value represents the mean \gls{pnl} that would be achieved under perfect foresight, considering all samples.

We observe that overall accuracy increases as we restrict the analysis to smaller sample shares across all periods.
Among the periods, period $1h\rightarrow\tfrac{h}{2}$ achieves the highest accuracy, followed by $3h\rightarrow2h$ and $2h\rightarrow1h$.
When comparing model accuracies, the gradient boosting model outperforms the logistic regression model in the periods $3h\rightarrow2h$ and $1h\rightarrow1$, whereas the logistic regression model outperforms the gradient booster during period $1h\rightarrow\tfrac{h}{2}$.
However, at very high sample shares, the accuracy declines slightly for both models during the periods $3h\rightarrow2h$ and $1h\rightarrow1$.
This reversal does not occur with respect to the \gls{pnl}.
At high signal strengths (at low sample share), the logistic regression model consistently outperforms the gradient booster in terms of \gls{pnl}.
The \gls{pnl} follows the same ordering across periods as the accuracy, with period $1h\rightarrow\tfrac{h}{2}$ yielding the highest values, followed by $3h\rightarrow2h$ and $2h\rightarrow1h$.
Interestingly, this ranking does not align with the mean \gls{pnl} under perfect foresight, indicated by the horizontal reference line in the figure.
Under perfect foresight, the mean \gls{pnl} increases across periods: from \euro{0.24} in period $3h\rightarrow2h$, to \euro{0.29} in period $2h\rightarrow1h$, and reaching \euro{0.48} in period $1h\rightarrow\tfrac{h}{2}$.

The increase in both accuracy and \gls{pnl} for samples with higher signal strength confirms that signal strength serves as a meaningful indicator of forecast correctness, which can be effectively leveraged in trading strategies.
Traders can use the signal strength by concentrating trading activity on forecasts with high signal strength, thereby improving overall performance.
The overall superior accuracy of the gradient boosting model indicates the model's ability to capture non-linear patterns.
However, the higher performance of the logistic regression over gradient boosting according to the \gls{pnl} metric likely results from the latter's tendency to overfit the training data. 
Logistic regression, by constraining the model to linear relationships, implicitly introduces regularization.
This regularizing constraint prevents the model from fitting non-linear patterns in the training data, thereby improving generalization to the test set and consequently the \gls{pnl}.
However, relying exclusively on linear models also introduces limitations.
The increasing potential \gls{pnl} across periods reflects rising market volatility, which aligns with expectations.
Nevertheless, the lower performance observed in the middle period ($2h\rightarrow1h$) is unexpected and may indicate greater market efficiency during this period.

\subsection{Temporal analysis}\label{sec:eval_temp}
\begin{figure*}[ht]
    \centering
    \includegraphics[width=\textwidth]{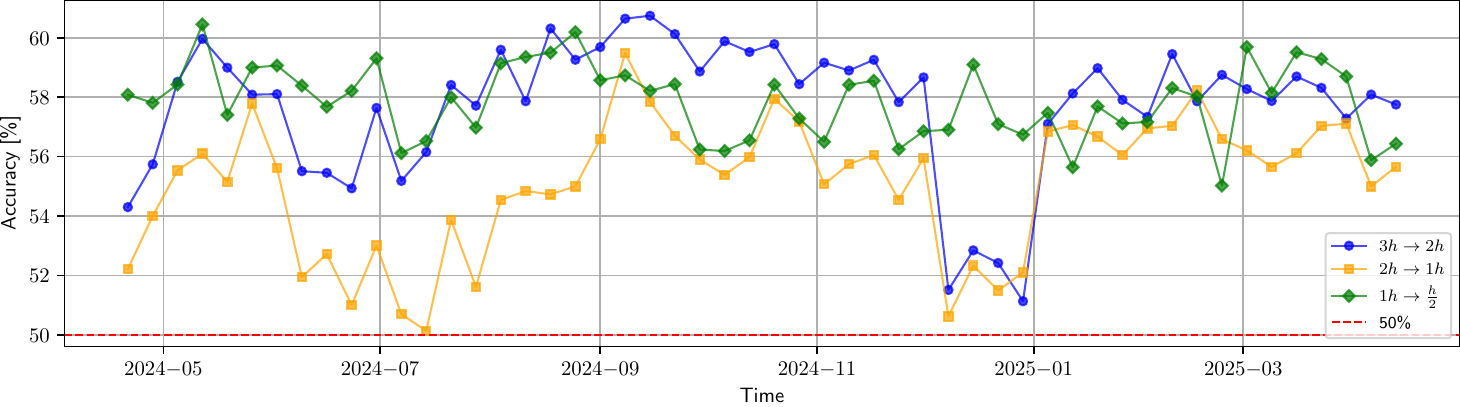}
    \caption{Weekly forecasting accuracy of the gradient boosting model using the best-performing feature set across the test interval.}
    \label{fig:acc_over_time}
\end{figure*}

To gain further insights into forecasting behavior, we analyze how forecasting accuracy evolves over time within the test interval. 
Figure~\ref{fig:acc_over_time} illustrates this by displaying weekly forecasting accuracy, where we compute a separate accuracy value for each week in the test set.

We observe that the accuracy for period $1h\rightarrow\tfrac{h}{2}$ remains relatively stable over time.
In contrast, periods $3h\rightarrow2h$ and $2h\rightarrow1h$ exhibit noticeable declines in accuracy around August 2024 and December 2024.
These drops are not present in period $1h\rightarrow\tfrac{h}{2}$.
Overall, period $2h\rightarrow1h$ yields the lowest accuracy, while periods $3h\rightarrow2h$ and $1h\rightarrow\tfrac{h}{2}$ exhibit the highest accuracy levels.

The declines observed in August and December 2024 for periods $3h\rightarrow2h$ and $2h\rightarrow1h$ may reflect episodes of elevated market volatility associated with high \gls{vre} penetration—mainly solar generation in summer and wind generation in winter.
These conditions can introduce irregularities in price dynamics, reducing the effectiveness of models trained on historical patterns.
In contrast, the stability of accuracy in period $1h\rightarrow\tfrac{h}{2}$ suggests that shorter lead time forecasts are less affected by such disruptions, likely due to their reliance on more immediate market indicators, such as features extracted from the \gls{lob}.

\subsection{Summary}\label{sec:eval_summary}
Our analysis spans three key dimensions: feature set comparison, model comparison, and temporal analysis. Among the features, those extracted from the \gls{lob} are the most informative for forecasting, followed by features derived from neighboring products. Notably, products with delivery starting before the current product contribute most to improving forecast accuracy.

Across the full test set, gradient boosting consistently outperforms logistic regression in terms of accuracy. However, the \gls{pnl} of the logistic regression is higher for samples with high signal strength.
This suggests that while logistic regression generalizes better overall, gradient boosting may capture more complex patterns and possibly overfit to the training data.

Signal strength serves as a reliable indicator of both forecast accuracy and trading profitability, as reflected in the \gls{pnl} metric. These results confirm a strong correlation between signal strength and directional accuracy, validating the probabilistic output of our model. As signal strength increases—i.e., as the model’s predicted probability deviates further from the 50\% threshold—forecast accuracy also improves. However, this comes at the cost of reduced sample coverage, as stricter signal strength thresholds exclude more data points.

Forecast performance also varies across time and forecasting periods. Period $1h\rightarrow\tfrac{h}{2}$ maintains relatively stable accuracy, indicating robustness against temporal fluctuations. In contrast, periods $3h\rightarrow2h$ and $2h\rightarrow1h$ exhibit noticeable drops in accuracy, particularly during the summer and winter months, likely due to increased market volatility associated with seasonal \gls{vre} dynamics.

From a practical perspective, these directional forecasts offer value in algorithmic trading applications. In speculative trading, traders can exploit short-term price movements by focusing on high-confidence forecasts. Since signal strength reflects forecast reliability, trading strategies can use it to manage risk—placing greater trust in signals with higher strength. Beyond speculative uses, these forecasts also assist in optimizing trade execution, for instance by improving the timing of position closures.

The robustness and high accuracy observed in the shortest lead-time period ($1h\rightarrow\tfrac{h}{2}$) highlight its particular suitability for trading strategies that rely on fast and reliable market signals.

\subsection{Limitations and Future Work}\label{sec:limitations}
Our study has several limitations.
We used logistic regression and gradient boosting as forecasting models; however, future work could explore additional \gls{ml} approaches, such as \glspl{nn}, to potentially enhance performance.

Moreover, our analysis focuses exclusively on the \gls{id3} trading period, which may limit the generalizability of our findings to other lead times.
Similarly, the test period spans 2024–2025, potentially limiting broader generalizability, as market behavior and volatility may vary over time.

We explore only a subset of features for price history and \glspl{lob}.
Future work could refine the feature design to further improve model accuracy.

In terms of feature selection, we only examined combinations involving the current product and one additional feature group.
This approach excludes more complex feature interactions and prevents an analysis of pair-wise dependencies that might improve forecasting performance. 
While our study concentrated on hourly products, future research could extend the method to quarter-hourly products to assess its broader applicability.

Finally, although our directional forecasts show promise, we did not evaluate their profitability within realistic trading strategies.
We chose to omit this step due to the inherent complexities of back-testing in the context of a \gls{lob}, which requires detailed assumptions about order execution, liquidity, and market impact.
Future work could address these challenges to assess the practical viability of implementing a strategy leveraging the forecasts in live trading environments.
\section{Conclusion}\label{sec:conclusion}
Short-term power trading, such as in the European \gls{cid} market, is becoming increasingly important due to the growing need to manage variable generation and consumption assets.
As trading activity in power markets accelerates, a growing number of participants are adopting algorithmic trading strategies.
\Gls{epf} tools can support these strategies by enhancing their decision-making capabilities.

In this study, we introduced a directional \gls{epf} model for the \gls{cid} market, focusing on hourly products and forecasting short-term market direction during the \gls{id3} trading period.
We investigated the impact of incorporating various exogenous variables—particularly price signals from neighboring products—on forecasting performance.

Our approach enables the integration of price information from related products into the forecasting model.
We find that including features derived from the \gls{lob} significantly improves forecast accuracy.
Moreover, incorporating price features from neighboring products with delivery periods overlapping the current trading period further enhances performance.
We also find that quarter-hourly products contribute positively to accuracy, even though they differ in temporal granularity.

In addition to producing directional forecasts, our method provides an estimate of the confidence in each prediction, enabling users to assess the reliability of the forecast.

Our results confirm that incorporating trading information from other \gls{cid} market products—especially those traded simultaneously with the target product—improves price forecasting accuracy.
Notably, we demonstrate that incorporating quarter-hourly products, an area not yet extensively explored in the literature, adds further value.
However, \gls{lob} features emerge as the most critical input, whereas fundamental and imbalance indicators offer limited improvement in predictive performance.

Overall, our directional forecasting method has practical relevance for the development of algorithmic trading strategies in the \gls{cid} market.

\begin{acks}
This research was funded in part by the Luxembourg National Research Fund (FNR) in the DELPHI Project, grant reference 17886330 / HPC BRIDGES/2022\_Phase2/17886330/DELPHI, FlexBEAM Project, grant reference 17742284, and by PayPal, PEARL grant reference 13342933 / Gilbert Fridgen.
For the purpose of open access, and in fulfillment of the obligations arising from the grant agreement, the author has applied a Creative Commons Attribution 4.0 International (CC BY 4.0) license to any Author Accepted Manuscript version arising from this submission.

The research was carried out as part of a partnership with the energy retailer Enovos Luxembourg S.A.

During the preparation of this work the authors used ChatGPT and Writefull in order to improve readability and language. After using this tool/service, the authors reviewed and edited the content as needed and take full responsibility for the content of the publication.

The authors thank Matthieu Sainlez and Sebastian Himpler for their valuable feedback and suggestions on this work.

\end{acks}

\bibliographystyle{ACM-Reference-Format}
\bibliography{99_references, 100_new_refs}


\begin{thebibliography}{54}


\ifx \showCODEN    \undefined \def \showCODEN     #1{\unskip}     \fi
\ifx \showDOI      \undefined \def \showDOI       #1{#1}\fi
\ifx \showISBNx    \undefined \def \showISBNx     #1{\unskip}     \fi
\ifx \showISBNxiii \undefined \def \showISBNxiii  #1{\unskip}     \fi
\ifx \showISSN     \undefined \def \showISSN      #1{\unskip}     \fi
\ifx \showLCCN     \undefined \def \showLCCN      #1{\unskip}     \fi
\ifx \shownote     \undefined \def \shownote      #1{#1}          \fi
\ifx \showarticletitle \undefined \def \showarticletitle #1{#1}   \fi
\ifx \showURL      \undefined \def \showURL       {\relax}        \fi
\providecommand\bibfield[2]{#2}
\providecommand\bibinfo[2]{#2}
\providecommand\natexlab[1]{#1}
\providecommand\showeprint[2][]{arXiv:#2}

\bibitem[{50Hertz Transmission GmbH} et~al\mbox{.}(2024)]%
        {50hertz_transmission_gmbh_nrv-saldo_2024}
\bibfield{author}{\bibinfo{person}{{50Hertz Transmission GmbH}}, \bibinfo{person}{{Amprion GmbH}}, \bibinfo{person}{{TenneT TSO GmbH}}, {and} \bibinfo{person}{{TransnetBW GmbH}}.} \bibinfo{year}{2024}\natexlab{}.
\newblock \bibinfo{title}{{NRV}-{Saldo}}.
\newblock
\newblock
\urldef\tempurl%
\url{https://www.netztransparenz.de/de-de/Regelenergie/NRV-und-RZ-Saldo/NRV-Saldo}
\showURL{%
\tempurl}


\bibitem[{All NEMO Committee}(2021)]%
        {all_nemo_committee_single_2021}
\bibfield{author}{\bibinfo{person}{{All NEMO Committee}}.} \bibinfo{year}{2021}\natexlab{}.
\newblock \bibinfo{booktitle}{\emph{Single {Intraday} {Coupling} ({XBID}) {Information} {Package}}}.
\newblock \bibinfo{type}{{T}echnical {R}eport}. \bibinfo{institution}{All NEMO Committee}.
\newblock
\urldef\tempurl%
\url{https://www.nemo-committee.eu/assets/files/SIDC_Information%20Package_April%202021-99076f6ed5001c4d47442ae5cccebf30.pdf}
\showURL{%
\tempurl}


\bibitem[Andrade et~al\mbox{.}(2017)]%
        {andrade_probabilistic_2017}
\bibfield{author}{\bibinfo{person}{José~R. Andrade}, \bibinfo{person}{Jorge Filipe}, \bibinfo{person}{Marisa Reis}, {and} \bibinfo{person}{Ricardo~J. Bessa}.} \bibinfo{year}{2017}\natexlab{}.
\newblock \showarticletitle{Probabilistic {Price} {Forecasting} for {Day}-{Ahead} and {Intraday} {Markets}: {Beyond} the {Statistical} {Model}}.
\newblock \bibinfo{journal}{\emph{Sustainability}} \bibinfo{volume}{9}, \bibinfo{number}{11} (\bibinfo{date}{Nov.} \bibinfo{year}{2017}), \bibinfo{pages}{1990}.
\newblock
\showISSN{2071-1050}
\urldef\tempurl%
\url{https://doi.org/10.3390/su9111990}
\showDOI{\tempurl}
\newblock
\shownote{Number: 11 Publisher: Multidisciplinary Digital Publishing Institute}.


\bibitem[Ayyadi et~al\mbox{.}(2019)]%
        {ayyadi2019optimal}
\bibfield{author}{\bibinfo{person}{Soumia Ayyadi}, \bibinfo{person}{Hasnae Bilil}, {and} \bibinfo{person}{Mohamed Maaroufi}.} \bibinfo{year}{2019}\natexlab{}.
\newblock \showarticletitle{Optimal charging of Electric Vehicles in residential area}.
\newblock \bibinfo{journal}{\emph{Sustainable Energy, Grids and Networks}}  \bibinfo{volume}{19} (\bibinfo{year}{2019}), \bibinfo{pages}{100240}.
\newblock


\bibitem[Baule and Naumann(2021)]%
        {baule_volatility_2021}
\bibfield{author}{\bibinfo{person}{Rainer Baule} {and} \bibinfo{person}{Michael Naumann}.} \bibinfo{year}{2021}\natexlab{}.
\newblock \showarticletitle{Volatility and {Dispersion} of {Hourly} {Electricity} {Contracts} on the {German} {Continuous} {Intraday} {Market}}.
\newblock \bibinfo{journal}{\emph{Energies}} \bibinfo{volume}{14}, \bibinfo{number}{22} (\bibinfo{date}{Jan.} \bibinfo{year}{2021}), \bibinfo{pages}{7531}.
\newblock
\showISSN{1996-1073}
\urldef\tempurl%
\url{https://doi.org/10.3390/en14227531}
\showDOI{\tempurl}
\newblock
\shownote{Number: 22 Publisher: Multidisciplinary Digital Publishing Institute}.


\bibitem[Birkeland and AlSkaif(2024)]%
        {birkeland_research_2024}
\bibfield{author}{\bibinfo{person}{Dane Birkeland} {and} \bibinfo{person}{Tarek AlSkaif}.} \bibinfo{year}{2024}\natexlab{}.
\newblock \showarticletitle{Research areas and methods of interest in {European} intraday electricity market research—{A} systematic literature review}.
\newblock \bibinfo{journal}{\emph{Sustainable Energy, Grids and Networks}}  \bibinfo{volume}{38} (\bibinfo{date}{June} \bibinfo{year}{2024}), \bibinfo{pages}{101368}.
\newblock
\showISSN{2352-4677}
\urldef\tempurl%
\url{https://doi.org/10.1016/j.segan.2024.101368}
\showDOI{\tempurl}


\bibitem[Ciarreta et~al\mbox{.}(2017)]%
        {ciarreta_modeling_2017}
\bibfield{author}{\bibinfo{person}{Aitor Ciarreta}, \bibinfo{person}{Peru Muniain}, {and} \bibinfo{person}{Ainhoa Zarraga}.} \bibinfo{year}{2017}\natexlab{}.
\newblock \showarticletitle{Modeling and forecasting realized volatility in {German}-{Austrian} continuous intraday electricity prices: {Journal} of {Forecasting}}.
\newblock \bibinfo{journal}{\emph{Journal of Forecasting}} \bibinfo{volume}{36}, \bibinfo{number}{6} (\bibinfo{date}{Sept.} \bibinfo{year}{2017}), \bibinfo{pages}{680--690}.
\newblock
\showISSN{02776693}
\urldef\tempurl%
\url{https://doi.org/10.1002/for.2463}
\showDOI{\tempurl}
\newblock
\shownote{Publisher: Wiley-Blackwell}.


\bibitem[Cramer et~al\mbox{.}(2023)]%
        {cramer_multivariate_2023}
\bibfield{author}{\bibinfo{person}{Eike Cramer}, \bibinfo{person}{Dirk Witthaut}, \bibinfo{person}{Alexander Mitsos}, {and} \bibinfo{person}{Manuel Dahmen}.} \bibinfo{year}{2023}\natexlab{}.
\newblock \showarticletitle{Multivariate probabilistic forecasting of intraday electricity prices using normalizing flows}.
\newblock \bibinfo{journal}{\emph{Applied Energy}}  \bibinfo{volume}{346} (\bibinfo{date}{Sept.} \bibinfo{year}{2023}), \bibinfo{pages}{121370}.
\newblock
\showISSN{0306-2619}
\urldef\tempurl%
\url{https://doi.org/10.1016/j.apenergy.2023.121370}
\showDOI{\tempurl}


\bibitem[Diebold and Mariano(1995)]%
        {diebold_comparing_1995}
\bibfield{author}{\bibinfo{person}{Francis Diebold} {and} \bibinfo{person}{Roberto Mariano}.} \bibinfo{year}{1995}\natexlab{}.
\newblock \showarticletitle{Comparing {Predictive} {Accuracy}}.
\newblock \bibinfo{journal}{\emph{Journal of Business \& Economic Statistics}} \bibinfo{volume}{13}, \bibinfo{number}{3} (\bibinfo{year}{1995}), \bibinfo{pages}{253--63}.
\newblock
\urldef\tempurl%
\url{https://econpapers.repec.org/article/besjnlbes/v_3a13_3ay_3a1995_3ai_3a3_3ap_3a253-63.htm}
\showURL{%
\tempurl}
\newblock
\shownote{Publisher: American Statistical Association}.


\bibitem[ENTSO-E(2025)]%
        {entso-e_entso-e_2025}
\bibfield{author}{\bibinfo{person}{ENTSO-E}.} \bibinfo{year}{2025}\natexlab{}.
\newblock \bibinfo{title}{{ENTSO}-{E} {Transparency} {Platform}}.
\newblock
\newblock
\urldef\tempurl%
\url{https://transparency.entsoe.eu/}
\showURL{%
\tempurl}


\bibitem[{EPEX SPOT SE}(2023)]%
        {epex_spot_se_description_2023}
\bibfield{author}{\bibinfo{person}{{EPEX SPOT SE}}.} \bibinfo{year}{2023}\natexlab{}.
\newblock \bibinfo{booktitle}{\emph{{DESCRIPTION} {OF} {EPEX} {SPOT} {MARKETS} {INDICES}}}.
\newblock \bibinfo{type}{{T}echnical {R}eport}. \bibinfo{institution}{EPEX SPOT SE}.
\newblock
\urldef\tempurl%
\url{https://www.epexspot.com/sites/default/files/download_center_files/EPEX%20SPOT%20Indices%202019-05_final.pdf}
\showURL{%
\tempurl}


\bibitem[{EPEX SPOT SE}(2024a)]%
        {epex_spot_se_market_2024}
\bibfield{author}{\bibinfo{person}{{EPEX SPOT SE}}.} \bibinfo{year}{2024}\natexlab{a}.
\newblock \bibinfo{title}{Market {Data} {\textbar} {EPEX} {SPOT}}.
\newblock
\newblock
\urldef\tempurl%
\url{https://www.epexspot.com/en/market-data}
\showURL{%
\tempurl}


\bibitem[{EPEX SPOT SE}(2024b)]%
        {epex_spot_se_new_2024}
\bibfield{author}{\bibinfo{person}{{EPEX SPOT SE}}.} \bibinfo{year}{2024}\natexlab{b}.
\newblock \bibinfo{title}{New 15-minute products in {Market} {Coupling} {\textbar} {EPEX} {SPOT}}.
\newblock
\newblock
\urldef\tempurl%
\url{https://www.epexspot.com/en/new-15-minute-products-market-coupling}
\showURL{%
\tempurl}


\bibitem[Frade et~al\mbox{.}(2018)]%
        {frade_influence_2018}
\bibfield{author}{\bibinfo{person}{Pedro M.~S. Frade}, \bibinfo{person}{João V. G.~A. Vieira-Costa}, \bibinfo{person}{Gerardo~J. Osório}, \bibinfo{person}{João J.~E. Santana}, {and} \bibinfo{person}{João P.~S. Catalão}.} \bibinfo{year}{2018}\natexlab{}.
\newblock \showarticletitle{Influence of {Wind} {Power} on {Intraday} {Electricity} {Spot} {Market}: {A} {Comparative} {Study} {Based} on {Real} {Data}}.
\newblock \bibinfo{journal}{\emph{Energies}} \bibinfo{volume}{11}, \bibinfo{number}{11} (\bibinfo{date}{Nov.} \bibinfo{year}{2018}), \bibinfo{pages}{2974}.
\newblock
\showISSN{1996-1073}
\urldef\tempurl%
\url{https://doi.org/10.3390/en11112974}
\showDOI{\tempurl}
\newblock
\shownote{Number: 11 Publisher: Multidisciplinary Digital Publishing Institute}.


\bibitem[Green et~al\mbox{.}(2006)]%
        {GREEN2006101}
\bibfield{author}{\bibinfo{person}{Bart~N. Green}, \bibinfo{person}{Claire~D. Johnson}, {and} \bibinfo{person}{Alan Adams}.} \bibinfo{year}{2006}\natexlab{}.
\newblock \showarticletitle{Writing narrative literature reviews for peer-reviewed journals: secrets of the trade}.
\newblock \bibinfo{journal}{\emph{Journal of Chiropractic Medicine}} \bibinfo{volume}{5}, \bibinfo{number}{3} (\bibinfo{year}{2006}), \bibinfo{pages}{101--117}.
\newblock
\showISSN{1556-3707}
\urldef\tempurl%
\url{https://doi.org/10.1016/S0899-3467(07)60142-6}
\showDOI{\tempurl}


\bibitem[Gürtler and Paulsen(2018)]%
        {gurtler_effect_2018}
\bibfield{author}{\bibinfo{person}{Marc Gürtler} {and} \bibinfo{person}{Thomas Paulsen}.} \bibinfo{year}{2018}\natexlab{}.
\newblock \showarticletitle{The effect of wind and solar power forecasts on day-ahead and intraday electricity prices in {Germany}}.
\newblock \bibinfo{journal}{\emph{Energy Economics}}  \bibinfo{volume}{75} (\bibinfo{date}{Sept.} \bibinfo{year}{2018}), \bibinfo{pages}{150--162}.
\newblock
\showISSN{0140-9883}
\urldef\tempurl%
\url{https://doi.org/10.1016/j.eneco.2018.07.006}
\showDOI{\tempurl}


\bibitem[Harvey et~al\mbox{.}(1997)]%
        {harvey_testing_1997}
\bibfield{author}{\bibinfo{person}{David Harvey}, \bibinfo{person}{Stephen Leybourne}, {and} \bibinfo{person}{Paul Newbold}.} \bibinfo{year}{1997}\natexlab{}.
\newblock \showarticletitle{Testing the equality of prediction mean squared errors}.
\newblock \bibinfo{journal}{\emph{International Journal of Forecasting}} \bibinfo{volume}{13}, \bibinfo{number}{2} (\bibinfo{date}{June} \bibinfo{year}{1997}), \bibinfo{pages}{281--291}.
\newblock
\showISSN{0169-2070}
\urldef\tempurl%
\url{https://doi.org/10.1016/S0169-2070(96)00719-4}
\showDOI{\tempurl}


\bibitem[Hastie(2009)]%
        {hastie2009elements}
\bibfield{author}{\bibinfo{person}{Trevor Hastie}.} \bibinfo{year}{2009}\natexlab{}.
\newblock \bibinfo{title}{The elements of statistical learning: data mining, inference, and prediction}.
\newblock
\newblock


\bibitem[Hirsch and Ziel(2024a)]%
        {hirsch_multivariate_2024}
\bibfield{author}{\bibinfo{person}{Simon Hirsch} {and} \bibinfo{person}{Florian Ziel}.} \bibinfo{year}{2024}\natexlab{a}.
\newblock \showarticletitle{Multivariate simulation-based forecasting for intraday power markets: {Modeling} cross-product price effects}.
\newblock \bibinfo{journal}{\emph{Applied Stochastic Models in Business and Industry}} \bibinfo{volume}{n/a}, \bibinfo{number}{n/a} (\bibinfo{year}{2024}).
\newblock
\showISSN{1526-4025}
\urldef\tempurl%
\url{https://doi.org/10.1002/asmb.2837}
\showDOI{\tempurl}
\newblock
\shownote{\_eprint: https://onlinelibrary.wiley.com/doi/pdf/10.1002/asmb.2837}.


\bibitem[Hirsch and Ziel(2024b)]%
        {hirsch_simulation-based_2024}
\bibfield{author}{\bibinfo{person}{Simon Hirsch} {and} \bibinfo{person}{Florian Ziel}.} \bibinfo{year}{2024}\natexlab{b}.
\newblock \showarticletitle{Simulation-based {Forecasting} for {Intraday} {Power} {Markets}: {Modelling} {Fundamental} {Drivers} for {Location}, {Shape} and {Scale} of the {Price} {Distribution}}.
\newblock \bibinfo{journal}{\emph{The Energy Journal}} \bibinfo{volume}{45}, \bibinfo{number}{3} (\bibinfo{date}{May} \bibinfo{year}{2024}), \bibinfo{pages}{107--144}.
\newblock
\showISSN{0195-6574, 1944-9089}
\urldef\tempurl%
\url{https://doi.org/10.5547/01956574.45.3.shir}
\showDOI{\tempurl}
\newblock
\shownote{arXiv:2211.13002 [econ, q-fin, stat]}.


\bibitem[Hirth and Mühlenpfordt(2021)]%
        {hirth_handel_2021}
\bibfield{author}{\bibinfo{person}{Lion Hirth} {and} \bibinfo{person}{Jonathan Mühlenpfordt}.} \bibinfo{year}{2021}\natexlab{}.
\newblock \bibinfo{booktitle}{\emph{Handel auf {Basis} des {Regelleistungs}-{Abrufs}}}.
\newblock \bibinfo{type}{{T}echnical {R}eport}. \bibinfo{institution}{Neon Neue Energieökonomik GmbH}.
\newblock
\urldef\tempurl%
\url{https://neon.energy/Neon_2021_Intraday_Regelleistung.pdf}
\showURL{%
\tempurl}


\bibitem[Hornek et~al\mbox{.}(2024)]%
        {hornek_comparative_2024}
\bibfield{author}{\bibinfo{person}{Timothée Hornek}, \bibinfo{person}{Sergio Potenciano~Menci}, \bibinfo{person}{Joaquí­n Delgado~Fernández}, {and} \bibinfo{person}{Ivan Pavić}.} \bibinfo{year}{2024}\natexlab{}.
\newblock \showarticletitle{Comparative {Analysis} of {Baseline} {Models} for {Rolling} {Price} {Forecasts} in the {German} {Continuous} {Intraday} {Electricity} {Market} {\textbar} {Energy} {Proceedings}}.
\newblock \bibinfo{journal}{\emph{Energy Proceedings}} \bibinfo{volume}{38}, \bibinfo{number}{Energy Transition towards Carbon Neutrality: Part I} (\bibinfo{date}{Jan.} \bibinfo{year}{2024}).
\newblock
\showISSN{2004-2965}
\urldef\tempurl%
\url{https://doi.org/10.46855/energy-proceedings-10885}
\showDOI{\tempurl}


\bibitem[{IRENA}(2024)]%
        {irena2024taxonomy}
\bibfield{author}{\bibinfo{person}{{IRENA}}.} \bibinfo{year}{2024}\natexlab{}.
\newblock \bibinfo{booktitle}{\emph{Energy Taxonomy: Classifications for the Energy Transition}}.
\newblock \bibinfo{type}{{T}echnical {R}eport}. \bibinfo{institution}{International Renewable Energy Agency}, \bibinfo{address}{Abu Dhabi}.
\newblock
\urldef\tempurl%
\url{https://www.irena.org/-/media/Files/IRENA/Agency/Publication/2024/Mar/IRENA_Energy_taxonomy_2024.pdf}
\showURL{%
\tempurl}
\newblock
\shownote{Accessed: 2025-05-08}.


\bibitem[Kath and Ziel(2018)]%
        {kath_value_2018}
\bibfield{author}{\bibinfo{person}{Christopher Kath} {and} \bibinfo{person}{Florian Ziel}.} \bibinfo{year}{2018}\natexlab{}.
\newblock \showarticletitle{The value of forecasts: {Quantifying} the economic gains of accurate quarter-hourly electricity price forecasts}.
\newblock \bibinfo{journal}{\emph{Energy Economics}}  \bibinfo{volume}{76} (\bibinfo{date}{Oct.} \bibinfo{year}{2018}), \bibinfo{pages}{411--423}.
\newblock
\showISSN{0140-9883}
\urldef\tempurl%
\url{https://doi.org/10.1016/j.eneco.2018.10.005}
\showDOI{\tempurl}


\bibitem[Ke et~al\mbox{.}(2017)]%
        {ke_lightgbm_2017}
\bibfield{author}{\bibinfo{person}{Guolin Ke}, \bibinfo{person}{Qi Meng}, \bibinfo{person}{Thomas Finley}, \bibinfo{person}{Taifeng Wang}, \bibinfo{person}{Wei Chen}, \bibinfo{person}{Weidong Ma}, \bibinfo{person}{Qiwei Ye}, {and} \bibinfo{person}{Tie-Yan Liu}.} \bibinfo{year}{2017}\natexlab{}.
\newblock \showarticletitle{{LightGBM}: {A} {Highly} {Efficient} {Gradient} {Boosting} {Decision} {Tree}}. In \bibinfo{booktitle}{\emph{Advances in {Neural} {Information} {Processing} {Systems}}}, Vol.~\bibinfo{volume}{30}. \bibinfo{publisher}{Curran Associates, Inc.}
\newblock
\urldef\tempurl%
\url{https://proceedings.neurips.cc/paper/2017/hash/6449f44a102fde848669bdd9eb6b76fa-Abstract.html}
\showURL{%
\tempurl}


\bibitem[Kiesel and Paraschiv(2017)]%
        {kiesel_econometric_2017}
\bibfield{author}{\bibinfo{person}{Rüdiger Kiesel} {and} \bibinfo{person}{Florentina Paraschiv}.} \bibinfo{year}{2017}\natexlab{}.
\newblock \showarticletitle{Econometric analysis of 15-minute intraday electricity prices}.
\newblock \bibinfo{journal}{\emph{Energy Economics}}  \bibinfo{volume}{64} (\bibinfo{date}{May} \bibinfo{year}{2017}), \bibinfo{pages}{77--90}.
\newblock
\showISSN{0140-9883}
\urldef\tempurl%
\url{https://doi.org/10.1016/j.eneco.2017.03.002}
\showDOI{\tempurl}


\bibitem[Koch and Hirth(2019)]%
        {koch_short-term_2019}
\bibfield{author}{\bibinfo{person}{Christopher Koch} {and} \bibinfo{person}{Lion Hirth}.} \bibinfo{year}{2019}\natexlab{}.
\newblock \showarticletitle{Short-term electricity trading for system balancing: {An} empirical analysis of the role of intraday trading in balancing {Germany}'s electricity system}.
\newblock \bibinfo{journal}{\emph{Renewable and Sustainable Energy Reviews}}  \bibinfo{volume}{113} (\bibinfo{date}{Oct.} \bibinfo{year}{2019}), \bibinfo{pages}{109275}.
\newblock
\showISSN{1364-0321}
\urldef\tempurl%
\url{https://doi.org/10.1016/j.rser.2019.109275}
\showDOI{\tempurl}


\bibitem[Koch and Maskos(2020)]%
        {koch_passive_2020}
\bibfield{author}{\bibinfo{person}{Christopher Koch} {and} \bibinfo{person}{Philipp Maskos}.} \bibinfo{year}{2020}\natexlab{}.
\newblock \showarticletitle{{PASSIVE} {BALANCING} {THROUGH} {INTRADAY} {TRADING}: {WHETHER} {INTERACTIONS} {BETWEEN} {SHORT}-{TERM} {TRADING} {AND} {BALANCING} {STABILIZE} {GERMANY}’{S} {ELECTRICITY} {SYSTEM}}.
\newblock \bibinfo{journal}{\emph{International Journal of Energy Economics and Policy}} \bibinfo{volume}{10}, \bibinfo{number}{1} (\bibinfo{date}{Jan.} \bibinfo{year}{2020}), \bibinfo{pages}{101--112}.
\newblock
\showISSN{21464553}
\urldef\tempurl%
\url{https://doi.org/10.32479/ijeep.8750}
\showDOI{\tempurl}


\bibitem[Kremer et~al\mbox{.}(2020a)]%
        {kremer_fundamental_2020}
\bibfield{author}{\bibinfo{person}{Marcel Kremer}, \bibinfo{person}{Rüdiger Kiesel}, {and} \bibinfo{person}{Florentina Paraschiv}.} \bibinfo{year}{2020}\natexlab{a}.
\newblock \showarticletitle{A {Fundamental} {Model} for {Continuous} {Intraday} {Electricity} {Trading}}.
\newblock \bibinfo{journal}{\emph{Philosophical Transactions of the Royal Society A: Mathematical, Physical and Engineering Sciences}} (\bibinfo{year}{2020}).
\newblock
\showISSN{1364-503X}
\urldef\tempurl%
\url{https://doi.org/10.2139/ssrn.3489214}
\showDOI{\tempurl}
\newblock
\shownote{Accepted: 2021-03-18T08:40:51Z Publisher: The Royal Society}.


\bibitem[Kremer et~al\mbox{.}(2020b)]%
        {kremer_intraday_2020}
\bibfield{author}{\bibinfo{person}{Marcel Kremer}, \bibinfo{person}{Rüdiger Kiesel}, {and} \bibinfo{person}{Florentina Paraschiv}.} \bibinfo{year}{2020}\natexlab{b}.
\newblock \showarticletitle{Intraday {Electricity} {Pricing} of {Night} {Contracts}}.
\newblock \bibinfo{journal}{\emph{Energies}} \bibinfo{volume}{13}, \bibinfo{number}{17} (\bibinfo{date}{Jan.} \bibinfo{year}{2020}), \bibinfo{pages}{4501}.
\newblock
\showISSN{1996-1073}
\urldef\tempurl%
\url{https://doi.org/10.3390/en13174501}
\showDOI{\tempurl}
\newblock
\shownote{Number: 17 Publisher: Multidisciplinary Digital Publishing Institute}.


\bibitem[{KU Leuven Energy Institute}(2015)]%
        {ku_leuven_energy_institute_ei_2015}
\bibfield{author}{\bibinfo{person}{{KU Leuven Energy Institute}}.} \bibinfo{year}{2015}\natexlab{}.
\newblock \bibinfo{booktitle}{\emph{{EI} {Fact} sheet: {The} current electricity market design in {Europe}}}.
\newblock \bibinfo{type}{{T}echnical {R}eport}. \bibinfo{institution}{KU Leuven Energy Institute}.
\newblock
\urldef\tempurl%
\url{https://set.kuleuven.be/ei/images/EI_factsheet8_eng.pdf/}
\showURL{%
\tempurl}


\bibitem[Kulakov and Ziel(2021)]%
        {kulakov_impact_2021}
\bibfield{author}{\bibinfo{person}{Sergei Kulakov} {and} \bibinfo{person}{Florian Ziel}.} \bibinfo{year}{2021}\natexlab{}.
\newblock \showarticletitle{The {Impact} of {Renewable} {Energy} {Forecasts} on {Intraday} {Electricity} {Prices}}.
\newblock \bibinfo{journal}{\emph{Economics of Energy \& Environmental Policy}} \bibinfo{volume}{10}, \bibinfo{number}{1} (\bibinfo{date}{Jan.} \bibinfo{year}{2021}).
\newblock
\showISSN{21605882}
\urldef\tempurl%
\url{https://doi.org/10.5547/2160-5890.10.1.skul}
\showDOI{\tempurl}
\newblock
\shownote{arXiv:1903.09641 [econ, q-fin]}.


\bibitem[Kuppelwieser and Wozabal(2023)]%
        {kuppelwieser_intraday_2023}
\bibfield{author}{\bibinfo{person}{Thomas Kuppelwieser} {and} \bibinfo{person}{David Wozabal}.} \bibinfo{year}{2023}\natexlab{}.
\newblock \showarticletitle{Intraday power trading: toward an arms race in weather forecasting?}
\newblock \bibinfo{journal}{\emph{OR Spectrum}} \bibinfo{volume}{45}, \bibinfo{number}{1} (\bibinfo{date}{March} \bibinfo{year}{2023}), \bibinfo{pages}{57--83}.
\newblock
\showISSN{1436-6304}
\urldef\tempurl%
\url{https://doi.org/10.1007/s00291-022-00698-5}
\showDOI{\tempurl}


\bibitem[Lago et~al\mbox{.}(2018)]%
        {lago_forecasting_2018}
\bibfield{author}{\bibinfo{person}{Jesus Lago}, \bibinfo{person}{Fjo De~Ridder}, {and} \bibinfo{person}{Bart De~Schutter}.} \bibinfo{year}{2018}\natexlab{}.
\newblock \showarticletitle{Forecasting spot electricity prices: {Deep} learning approaches and empirical comparison of traditional algorithms}.
\newblock \bibinfo{journal}{\emph{Applied Energy}}  \bibinfo{volume}{221} (\bibinfo{date}{July} \bibinfo{year}{2018}), \bibinfo{pages}{386--405}.
\newblock
\showISSN{0306-2619}
\urldef\tempurl%
\url{https://doi.org/10.1016/j.apenergy.2018.02.069}
\showDOI{\tempurl}


\bibitem[Maciejowska et~al\mbox{.}(2019)]%
        {maciejowska_day-ahead_2019}
\bibfield{author}{\bibinfo{person}{Katarzyna Maciejowska}, \bibinfo{person}{Weronika Nitka}, {and} \bibinfo{person}{Tomasz Weron}.} \bibinfo{year}{2019}\natexlab{}.
\newblock \showarticletitle{Day-{Ahead} vs. {Intraday}—{Forecasting} the {Price} {Spread} to {Maximize} {Economic} {Benefits}}.
\newblock \bibinfo{journal}{\emph{Energies}} \bibinfo{volume}{12}, \bibinfo{number}{4} (\bibinfo{date}{Jan.} \bibinfo{year}{2019}), \bibinfo{pages}{631}.
\newblock
\showISSN{1996-1073}
\urldef\tempurl%
\url{https://doi.org/10.3390/en12040631}
\showDOI{\tempurl}
\newblock
\shownote{Number: 4 Publisher: Multidisciplinary Digital Publishing Institute}.


\bibitem[Maciejowska et~al\mbox{.}(2020)]%
        {maciejowska_pca_2020}
\bibfield{author}{\bibinfo{person}{Katarzyna Maciejowska}, \bibinfo{person}{Bartosz Uniejewski}, {and} \bibinfo{person}{Tomasz Serafin}.} \bibinfo{year}{2020}\natexlab{}.
\newblock \showarticletitle{{PCA} {Forecast} {Averaging}—{Predicting} {Day}-{Ahead} and {Intraday} {Electricity} {Prices}}.
\newblock \bibinfo{journal}{\emph{Energies}} \bibinfo{volume}{13}, \bibinfo{number}{14} (\bibinfo{date}{Jan.} \bibinfo{year}{2020}), \bibinfo{pages}{3530}.
\newblock
\showISSN{1996-1073}
\urldef\tempurl%
\url{https://doi.org/10.3390/en13143530}
\showDOI{\tempurl}
\newblock
\shownote{Number: 14 Publisher: Multidisciplinary Digital Publishing Institute}.


\bibitem[Marcjasz et~al\mbox{.}(2020)]%
        {marcjasz_beating_2020}
\bibfield{author}{\bibinfo{person}{Grzegorz Marcjasz}, \bibinfo{person}{Bartosz Uniejewski}, {and} \bibinfo{person}{Rafał Weron}.} \bibinfo{year}{2020}\natexlab{}.
\newblock \showarticletitle{Beating the {Naïve}—{Combining} {LASSO} with {Naïve} {Intraday} {Electricity} {Price} {Forecasts}}.
\newblock \bibinfo{journal}{\emph{Energies}} \bibinfo{volume}{13}, \bibinfo{number}{7} (\bibinfo{date}{Jan.} \bibinfo{year}{2020}), \bibinfo{pages}{1667}.
\newblock
\showISSN{1996-1073}
\urldef\tempurl%
\url{https://doi.org/10.3390/en13071667}
\showDOI{\tempurl}
\newblock
\shownote{Number: 7 Publisher: Multidisciplinary Digital Publishing Institute}.


\bibitem[{Market Coupling Steering Committee}(2023)]%
        {market_coupling_steering_committee_sidc_2023}
\bibfield{author}{\bibinfo{person}{{Market Coupling Steering Committee}}.} \bibinfo{year}{2023}\natexlab{}.
\newblock \bibinfo{booktitle}{\emph{{SIDC} {Stakeholder} {Report} {October} 2023}}.
\newblock \bibinfo{type}{{T}echnical {R}eport}. \bibinfo{institution}{Market Coupling Steering Committee}.
\newblock
\urldef\tempurl%
\url{https://www.nemo-committee.eu/assets/files/SIDC_stakeholder_report_1023-f284194310bdf658e82e27220a285de1.pdf}
\showURL{%
\tempurl}


\bibitem[Meeus(2020)]%
        {meeus_evolution_2020}
\bibfield{author}{\bibinfo{person}{Leonardo Meeus}.} \bibinfo{year}{2020}\natexlab{}.
\newblock \bibinfo{booktitle}{\emph{The {Evolution} of {Electricity} {Markets} in {Europe}}}.
\newblock
\showISBNx{978-1-78990-547-2}
\urldef\tempurl%
\url{https://doi.org/10.4337/9781789905472}
\showDOI{\tempurl}


\bibitem[Narajewski and Ziel(2020a)]%
        {narajewski_econometric_2020}
\bibfield{author}{\bibinfo{person}{Michał Narajewski} {and} \bibinfo{person}{Florian Ziel}.} \bibinfo{year}{2020}\natexlab{a}.
\newblock \showarticletitle{Econometric modelling and forecasting of intraday electricity prices}.
\newblock \bibinfo{journal}{\emph{Journal of Commodity Markets}}  \bibinfo{volume}{19} (\bibinfo{date}{Sept.} \bibinfo{year}{2020}), \bibinfo{pages}{100107}.
\newblock
\showISSN{2405-8513}
\urldef\tempurl%
\url{https://doi.org/10.1016/j.jcomm.2019.100107}
\showDOI{\tempurl}


\bibitem[Narajewski and Ziel(2020b)]%
        {narajewski_ensemble_2020}
\bibfield{author}{\bibinfo{person}{Michał Narajewski} {and} \bibinfo{person}{Florian Ziel}.} \bibinfo{year}{2020}\natexlab{b}.
\newblock \showarticletitle{Ensemble forecasting for intraday electricity prices: {Simulating} trajectories}.
\newblock \bibinfo{journal}{\emph{Applied Energy}}  \bibinfo{volume}{279} (\bibinfo{date}{Dec.} \bibinfo{year}{2020}), \bibinfo{pages}{115801}.
\newblock
\showISSN{0306-2619}
\urldef\tempurl%
\url{https://doi.org/10.1016/j.apenergy.2020.115801}
\showDOI{\tempurl}


\bibitem[Paré et~al\mbox{.}(2015)]%
        {pare_synthesizing_2015}
\bibfield{author}{\bibinfo{person}{Guy Paré}, \bibinfo{person}{Marie-Claude Trudel}, \bibinfo{person}{Mirou Jaana}, {and} \bibinfo{person}{Spyros Kitsiou}.} \bibinfo{year}{2015}\natexlab{}.
\newblock \showarticletitle{Synthesizing information systems knowledge: {A} typology of literature reviews}.
\newblock \bibinfo{journal}{\emph{Information \& Management}} \bibinfo{volume}{52}, \bibinfo{number}{2} (\bibinfo{date}{March} \bibinfo{year}{2015}), \bibinfo{pages}{183--199}.
\newblock
\showISSN{0378-7206}
\urldef\tempurl%
\url{https://doi.org/10.1016/j.im.2014.08.008}
\showDOI{\tempurl}


\bibitem[Pedregosa et~al\mbox{.}(2011)]%
        {pedregosa_scikit-learn_2011}
\bibfield{author}{\bibinfo{person}{Fabian Pedregosa}, \bibinfo{person}{Gaël Varoquaux}, \bibinfo{person}{Alexandre Gramfort}, \bibinfo{person}{Vincent Michel}, \bibinfo{person}{Bertrand Thirion}, \bibinfo{person}{Olivier Grisel}, \bibinfo{person}{Mathieu Blondel}, \bibinfo{person}{Peter Prettenhofer}, \bibinfo{person}{Ron Weiss}, \bibinfo{person}{Vincent Dubourg}, \bibinfo{person}{Jake Vanderplas}, \bibinfo{person}{Alexandre Passos}, \bibinfo{person}{David Cournapeau}, \bibinfo{person}{Matthieu Brucher}, \bibinfo{person}{Matthieu Perrot}, {and} \bibinfo{person}{Edouard Duchesnay}.} \bibinfo{year}{2011}\natexlab{}.
\newblock \showarticletitle{Scikit-learn: {Machine} {Learning} in {Python}}.
\newblock \bibinfo{journal}{\emph{Journal of Machine Learning Research}} \bibinfo{volume}{12}, \bibinfo{number}{85} (\bibinfo{year}{2011}), \bibinfo{pages}{2825--2830}.
\newblock
\showISSN{1533-7928}
\urldef\tempurl%
\url{http://jmlr.org/papers/v12/pedregosa11a.html}
\showURL{%
\tempurl}


\bibitem[Peffers et~al\mbox{.}(2007)]%
        {peffers_design_2007}
\bibfield{author}{\bibinfo{person}{Ken Peffers}, \bibinfo{person}{Tuure Tuunanen}, \bibinfo{person}{Marcus~A Rothenberger}, {and} \bibinfo{person}{Samir Chatterjee}.} \bibinfo{year}{2007}\natexlab{}.
\newblock \showarticletitle{A design science research methodology for information systems research}.
\newblock \bibinfo{journal}{\emph{Journal of management information systems}} \bibinfo{volume}{24}, \bibinfo{number}{3} (\bibinfo{year}{2007}), \bibinfo{pages}{45--77}.
\newblock
\newblock
\shownote{Publisher: Taylor \& Francis}.


\bibitem[{Regelleistung}(2025)]%
        {regelleistung_gcc}
\bibfield{author}{\bibinfo{person}{{Regelleistung}}.} \bibinfo{year}{2025}\natexlab{}.
\newblock \bibinfo{title}{German Grid Control Cooperation}.
\newblock \bibinfo{howpublished}{\url{https://www.regelleistung.net/en-us/Market-information/German-grid-control-cooperation}}.
\newblock
\newblock
\shownote{Accessed: 2025-05-09}.


\bibitem[Scholz et~al\mbox{.}(2021)]%
        {scholz_towards_2021}
\bibfield{author}{\bibinfo{person}{Christoph Scholz}, \bibinfo{person}{Malte Lehna}, \bibinfo{person}{Katharina Brauns}, {and} \bibinfo{person}{André Baier}.} \bibinfo{year}{2021}\natexlab{}.
\newblock \showarticletitle{Towards the {Prediction} of {Electricity} {Prices} at the {Intraday} {Market} {Using} {Shallow} and {Deep}-{Learning} {Methods}}. In \bibinfo{booktitle}{\emph{Mining {Data} for {Financial} {Applications}}} \emph{(\bibinfo{series}{Lecture {Notes} in {Computer} {Science}})}, \bibfield{editor}{\bibinfo{person}{Valerio Bitetta}, \bibinfo{person}{Ilaria Bordino}, \bibinfo{person}{Andrea Ferretti}, \bibinfo{person}{Francesco Gullo}, \bibinfo{person}{Giovanni Ponti}, {and} \bibinfo{person}{Lorenzo Severini}} (Eds.). \bibinfo{publisher}{Springer International Publishing}, \bibinfo{address}{Cham}, \bibinfo{pages}{101--118}.
\newblock
\showISBNx{978-3-030-66981-2}
\urldef\tempurl%
\url{https://doi.org/10.1007/978-3-030-66981-2_9}
\showDOI{\tempurl}


\bibitem[Serafin et~al\mbox{.}(2022)]%
        {serafin_trading_2022}
\bibfield{author}{\bibinfo{person}{Tomasz Serafin}, \bibinfo{person}{Grzegorz Marcjasz}, {and} \bibinfo{person}{Rafał Weron}.} \bibinfo{year}{2022}\natexlab{}.
\newblock \showarticletitle{Trading on short-term path forecasts of intraday electricity prices}.
\newblock \bibinfo{journal}{\emph{Energy Economics}}  \bibinfo{volume}{112} (\bibinfo{date}{Aug.} \bibinfo{year}{2022}), \bibinfo{pages}{106125}.
\newblock
\showISSN{0140-9883}
\urldef\tempurl%
\url{https://doi.org/10.1016/j.eneco.2022.106125}
\showDOI{\tempurl}


\bibitem[Shinde and Amelin(2019)]%
        {shinde_literature_2019}
\bibfield{author}{\bibinfo{person}{Priyanka Shinde} {and} \bibinfo{person}{Mikael Amelin}.} \bibinfo{year}{2019}\natexlab{}.
\newblock \showarticletitle{A {Literature} {Review} of {Intraday} {Electricity} {Markets} and {Prices}}. In \bibinfo{booktitle}{\emph{2019 {IEEE} {Milan} {PowerTech}}}. \bibinfo{pages}{1--6}.
\newblock
\urldef\tempurl%
\url{https://doi.org/10.1109/PTC.2019.8810752}
\showDOI{\tempurl}


\bibitem[Singh and Singh(2020)]%
        {SINGH2020105524}
\bibfield{author}{\bibinfo{person}{Dalwinder Singh} {and} \bibinfo{person}{Birmohan Singh}.} \bibinfo{year}{2020}\natexlab{}.
\newblock \showarticletitle{Investigating the impact of data normalization on classification performance}.
\newblock \bibinfo{journal}{\emph{Applied Soft Computing}}  \bibinfo{volume}{97} (\bibinfo{year}{2020}), \bibinfo{pages}{105524}.
\newblock
\showISSN{1568-4946}
\urldef\tempurl%
\url{https://doi.org/10.1016/j.asoc.2019.105524}
\showDOI{\tempurl}


\bibitem[{The Netherlands Authority for Consumers {and} Markets}(2024)]%
        {the_netherlands_authority_for_consumers_and_markets_algorithmic_2024}
\bibfield{author}{\bibinfo{person}{{The Netherlands Authority for Consumers {and} Markets}}.} \bibinfo{year}{2024}\natexlab{}.
\newblock \bibinfo{booktitle}{\emph{Algorithmic {Trading} in {Wholesale} {Energy} {Markets}}}.
\newblock \bibinfo{type}{{T}echnical {R}eport}.
\newblock
\urldef\tempurl%
\url{https://www.acm.nl/system/files/documents/rapport-marktstudie-algoritmische-handel-energiemarkt-public_0.pdf}
\showURL{%
\tempurl}


\bibitem[Uniejewski et~al\mbox{.}(2019)]%
        {uniejewski_understanding_2019}
\bibfield{author}{\bibinfo{person}{Bartosz Uniejewski}, \bibinfo{person}{Grzegorz Marcjasz}, {and} \bibinfo{person}{Rafał Weron}.} \bibinfo{year}{2019}\natexlab{}.
\newblock \showarticletitle{Understanding intraday electricity markets: {Variable} selection and very short-term price forecasting using {LASSO}}.
\newblock \bibinfo{journal}{\emph{International Journal of Forecasting}} \bibinfo{volume}{35}, \bibinfo{number}{4} (\bibinfo{date}{Oct.} \bibinfo{year}{2019}), \bibinfo{pages}{1533--1547}.
\newblock
\showISSN{0169-2070}
\urldef\tempurl%
\url{https://doi.org/10.1016/j.ijforecast.2019.02.001}
\showDOI{\tempurl}


\bibitem[Uniejewski et~al\mbox{.}(2018)]%
        {uniejewski_variance_2018}
\bibfield{author}{\bibinfo{person}{Bartosz Uniejewski}, \bibinfo{person}{Rafał Weron}, {and} \bibinfo{person}{Florian Ziel}.} \bibinfo{year}{2018}\natexlab{}.
\newblock \showarticletitle{Variance {Stabilizing} {Transformations} for {Electricity} {Spot} {Price} {Forecasting}}.
\newblock \bibinfo{journal}{\emph{IEEE Transactions on Power Systems}} \bibinfo{volume}{33}, \bibinfo{number}{2} (\bibinfo{date}{March} \bibinfo{year}{2018}), \bibinfo{pages}{2219--2229}.
\newblock
\showISSN{1558-0679}
\urldef\tempurl%
\url{https://doi.org/10.1109/TPWRS.2017.2734563}
\showDOI{\tempurl}
\newblock
\shownote{Conference Name: IEEE Transactions on Power Systems}.


\bibitem[Zachmann et~al\mbox{.}(2023)]%
        {zachmann_design_2023}
\bibfield{author}{\bibinfo{person}{Georg Zachmann}, \bibinfo{person}{Lion Hirth}, \bibinfo{person}{Conall Heussaff}, \bibinfo{person}{Ingmar Schlecht}, \bibinfo{person}{Jonathan Mühlenpfordt}, {and} \bibinfo{person}{Anselm Eicke}.} \bibinfo{year}{2023}\natexlab{}.
\newblock \bibinfo{booktitle}{\emph{The design of the {European} electricity market - {Current} proposals and ways ahead}}.
\newblock \bibinfo{type}{{T}echnical {R}eport}. \bibinfo{institution}{Policy Department for Economic, Scientific and Quality of Life Policies Directorate-General for Internal Policies}.
\newblock
\urldef\tempurl%
\url{https://www.europarl.europa.eu/RegData/etudes/STUD/2023/740094/IPOL_STU(2023)740094_EN.pdf}
\showURL{%
\tempurl}


\bibitem[Ziel(2017)]%
        {ziel_modeling_2017}
\bibfield{author}{\bibinfo{person}{Florian Ziel}.} \bibinfo{year}{2017}\natexlab{}.
\newblock \showarticletitle{Modeling the impact of wind and solar power forecasting errors on intraday electricity prices}. In \bibinfo{booktitle}{\emph{2017 14th {International} {Conference} on the {European} {Energy} {Market} ({EEM})}}. \bibinfo{publisher}{IEEE}, \bibinfo{address}{Dresden, Germany}, \bibinfo{pages}{1--5}.
\newblock
\showISBNx{978-1-5090-5499-2}
\urldef\tempurl%
\url{https://doi.org/10.1109/EEM.2017.7981900}
\showDOI{\tempurl}


\end{thebibliography}

\appendix
\section{\gls{cid} Price Features}\label{app:cid_price_ftr}
\subsection{\gls{cid} Notation}\label{app:cid_notation}
We introduce a notation for \gls{cid} transactions, drawing on the conventions established by~\citet{narajewski_econometric_2020} and incorporating elements from~\citet{hornek_comparative_2024}.
We describe \gls{cid} products as tuples \gls{product}, with \gls{dlvrystart} marking the start of product delivery and \gls{prodlen} defining the length of the product.
For hourly products, \gls{prodlen} is set to 1 hour, and for quarter-hourly products, it is 15 minutes. We indicate the delivery period of a product by the interval $[s, s+\ell)$.

Furthermore, \gls{prodtrades} encompasses the set of trades within the product \gls{product}.
We characterize each trade $k \in \gls{prodtrades}$ by its execution time (\gls{tradetime}), trade volume (\gls{tradevol} in $MW$), and trade price (\gls{tradeprice} in \euro/$MW$).

Using this notation, we define the \gls{vwap} and \gls{vwsd} for any subset of trades, $\gls{tradesubset} \subseteq \gls{prodtrades}$, as specified in Equations~(\ref{eq:vwap}) and~(\ref{eq:vwsd}).
\begin{align}
    \text{\gls{vwap}}&(\gls{tradesubset}):=\frac{1}{\sum_{k\in\gls{tradesubset}}\gls{tradevol}}\sum_{k\in\mathcal{S}^{s,\ell}}{\gls{tradevol}\gls{tradeprice}}\label{eq:vwap}\\
    \text{\gls{vwsd}}&(\gls{tradesubset}):=\sqrt{\frac{\sum_{k\in\gls{tradesubset}}{\gls{tradevol}(\gls{tradeprice}-\text{\gls{vwap}}(\gls{tradesubset}))^2}}{\frac{|\gls{tradesubset}|-1}{|\gls{tradesubset}|}\sum_{k\in\gls{tradesubset}}\gls{tradevol}}}\label{eq:vwsd}
\end{align}

\subsection{\gls{cid} Price Feature Computation}\label{app:cid_price_ftr_computation}
This section introduces the methodology for computing price features, which is independent of product length. 
Let \gls{curtime} denote the current time.
At current time (\gls{curtime}), the \gls{vwap} of of the four most recent trades is calculated as shown in Equation~(\ref{eq:curprice}).
The set $\mathcal{S}^{s,\ell}_{\text{last4},t}$ includes the four trades with the most recent execution time (\gls{tradetime}) within the subset $\{k\in\gls{prodtrades}:\gls{tradetime}\leq\gls{curtime}\}$, serving as an estimate for the current market price~\cite{hornek_comparative_2024}:
\begin{equation}
    \gls{curprice}:=\text{\gls{vwap}}(\mathcal{S}^{s,\ell}_{\text{last4},t})\label{eq:curprice}
\end{equation}
Furthermore, we define the \gls{vwap} of trades within a past interval as detailed in Equation~(\ref{eq:lagprice}).
The interval's length is denoted by $\Delta>0$ and the lag by $h\in\mathbb{N}_0$.
\begin{equation}
    \gls{lagprice}:=\text{\gls{vwap}}(\{k\in\gls{prodtrades}:\gls{curtime}-(h+1)\Delta\leq\gls{tradetime}\leq\gls{curtime}-h\Delta\})\label{eq:lagprice}
\end{equation}
We define \gls{vwsd} of trades within a past interval (\gls{lagpricesd}) analogously.
Assuming the the maximum number of lags is $h_{\max}\in\mathbb{N}$, we define the historical price vector (\gls{priceftr}) in Equation~(\ref{eq:priceftr}).
\begin{equation}
    \gls{priceftr}:=(\text{VWAP}^{s,\ell}_{t,0,\Delta},\text{VWAP}^{s,\ell}_{t,1,\Delta},...,\text{VWAP}^{s,\ell}_{t,h_{\max}-1,\Delta})^T\label{eq:priceftr}
\end{equation}

\section{Price Feature Transformation}\label{app:price_ftr_norm}
The following details the price normalization procedure we follow, building on notation introduced in~\ref{app:cid_price_ftr}.
Let $x$ be an arbitrary price feature used to forecast a product \gls{product} at time $t$.
We transform price features in two steps.
We normalize the price data, as expressed in Equation~(\ref{eq:price_ftr_norm}), where $h=0$ and $\Delta=\infty$, meaning that we take all historical trades that happened before the current time (\gls{curtime}) into account to estimate the mean (\gls{lagprice}) and standard deviation (\gls{lagpricesd})~\cite{hornek_comparative_2024}.
\begin{equation}
    z:=\frac{x-\gls{lagprice}}{\gls{lagpricesd}}\label{eq:price_ftr_norm}
\end{equation}

\section{\gls{dm} Testing for \gls{cid} \gls{epf}}\label{app:dm_testing_cid_epf}
The \gls{dm} test is a statistical method used to compare the forecasting accuracy of different models~\cite{diebold_comparing_1995}.
It requires time-series of forecasting errors to compute the test statistic.
However, the multiple price forecasting approach to \gls{epf} in the \gls{cid} market produces a series of errors for each product, resulting in multiple error series rather than a single error series.

To address this, following \citet{hirsch_simulation-based_2024}, we aggregate errors on a per-product basis for each forecasting interval ($3h\rightarrow2h$, $2h\rightarrow1h$, and $1h\rightarrow\tfrac{h}{2}$; see Section~\ref{sec:neighboring_products}). For each product, we calculate the error rate (1 minus accuracy), creating a time series of error metrics across sequential products. We then compute the difference between the error series of the two models, yielding a series of error differences appropriate for statistical testing.

Before applying the \gls{dm} test, we assess the stationarity of the error difference series using the \gls{adf} test at a 0.01 significance level, as stationarity is a key assumption of the \gls{dm} test. Once confirmed, we apply the \gls{dm} test using the modified version proposed by \citet{harvey_testing_1997}, with a significance level of 0.05.

\section{\Acrlong{dm} Test Results}\label{app:dm_test_results}

\begin{figure*}[htbp!]
  \centering
  
  \begin{subfigure}[b]{\columnwidth}
    \centering
    \includegraphics[width=\linewidth]{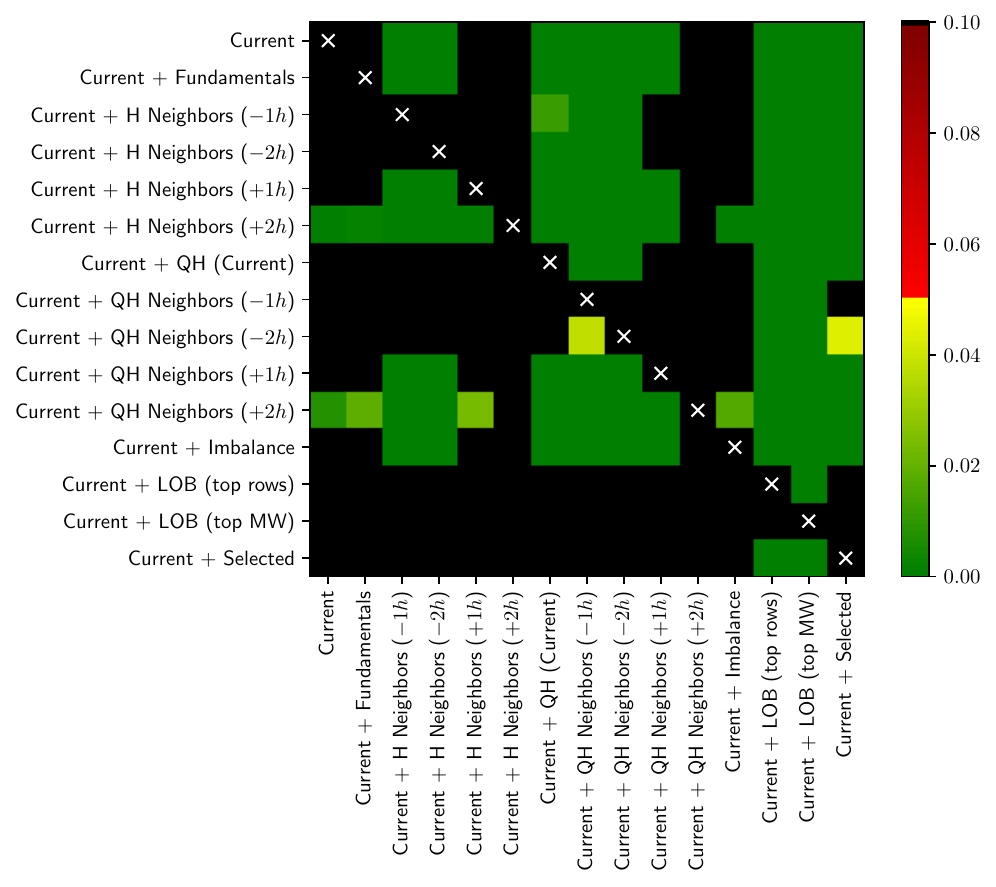}
    \caption{Period $3\rightarrow2$}
    \label{fig:dm_period_a}
  \end{subfigure}
  \hfill
  \begin{subfigure}[b]{\columnwidth}
    \centering
    \includegraphics[width=\linewidth]{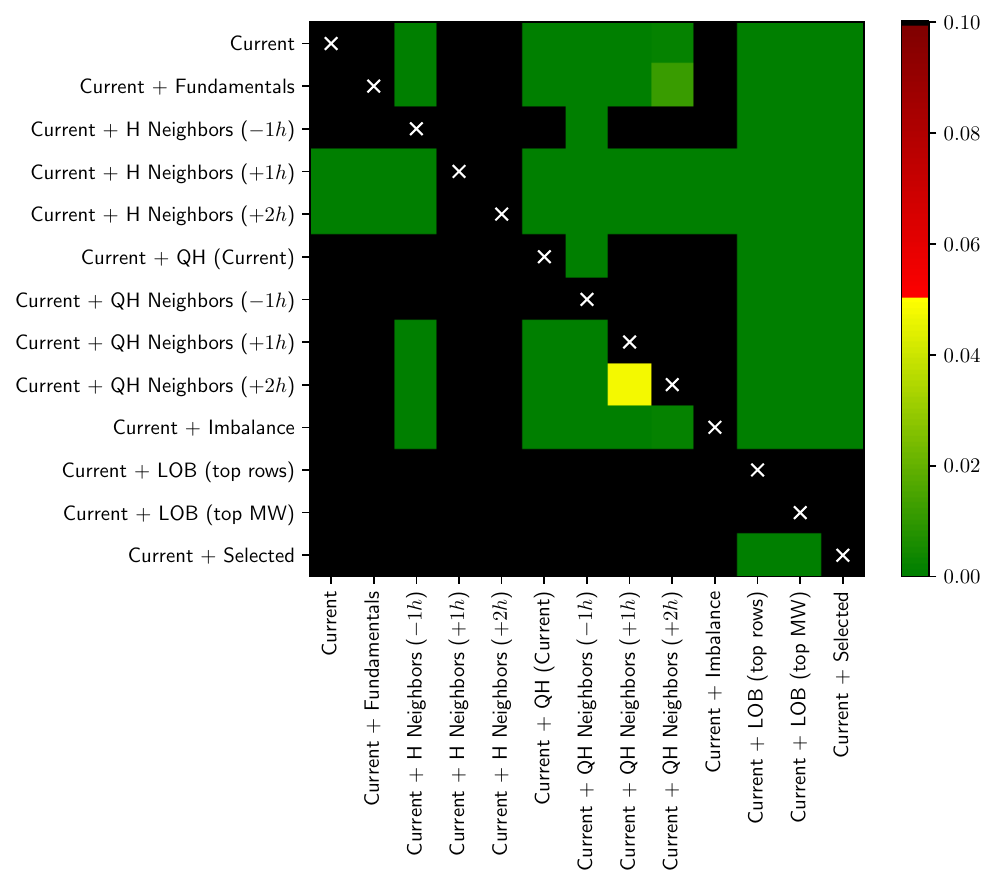}
    \caption{Period $2\rightarrow1$}
    \label{fig:dm_period_b}
  \end{subfigure}
  \hfill
  \vspace{1.0cm}
  \begin{subfigure}[b]{\columnwidth}
    \centering
    \includegraphics[width=\linewidth]{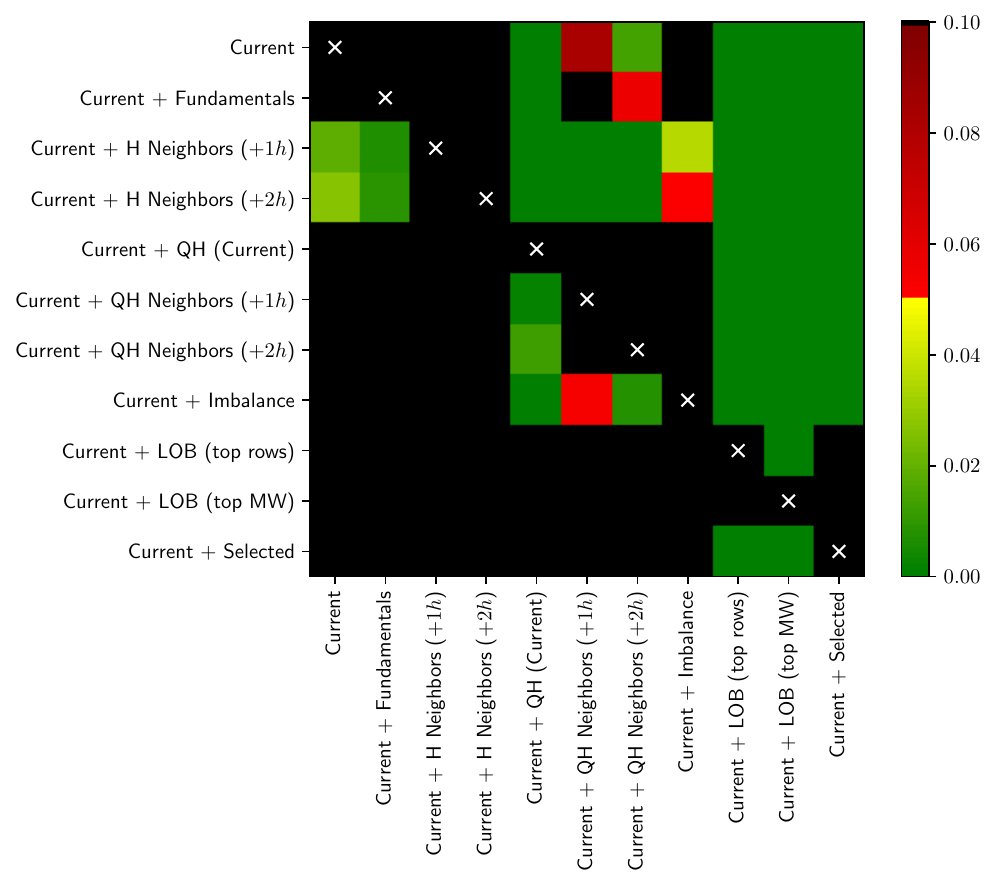}
    \caption{Period $1\rightarrow\tfrac{1}{2}$}
    \label{fig:dm_period_c}
  \end{subfigure}
  \caption{\Gls{dm} test results for the gradient boosting model accuracies across periods.}
  \label{fig:dm_period_a_b_c}
\end{figure*}

Figure~\ref{fig:dm_period_a_b_c} presents additional \gls{dm} test results, comparing the accuracy of the gradient boosting models using different feature sets. The chessboard-shaped heat map displays p-values indicating whether the forecast on the y-axis is significantly more accurate than the x-axis forecast; values near 0 suggest the x-axis forecast is significantly better~\cite{lago_forecasting_2018}.



\end{document}